\documentclass[12pt]{iopart_mod}

\usepackage{amsmath}
\usepackage{amsfonts}
\usepackage{mathrsfs,cite}

\usepackage{tikz}
\usetikzlibrary{shapes,arrows,shadows}
\usepackage{tangocolors}
\usepackage{afterpage}

\newcommand{\beqs}{\begin{equation*}}
\newcommand{\beq}{\begin{equation}}

\newcommand{\eeqs}{\end{equation*}}
\newcommand{\eeq}{\end{equation}}

\newcommand{\beqas}{\begin{eqnarray*}}
\newcommand{\beqa}{\begin{eqnarray}}

\newcommand{\eeqas}{\end{eqnarray*}}
\newcommand{\eeqa}{\end{eqnarray}}

\newcommand{\eq}[2]{\begin{equation} #1 \label{#2} \end{equation}}

\newcommand{\eps}{\varepsilon}
\newcommand{\al}{\alpha}
\newcommand{\be}{\beta}

\newcommand{\de}{\delta}

\newcommand{\la}{\lambda}
\newcommand{\si}{\sigma}

\newcommand{\Ga}{\Gamma}

\newcommand{\La}{\Lambda}

\newcommand{\blist}{\begin{itemize}}

\newcommand{\elist}{\end{itemize}}

\providecommand{\href}[2]{#2}

\DeclareFontFamily{OT1}{rsfs}{}
\DeclareFontShape{OT1}{rsfs}{m}{n}{ <-7> rsfs5 <7-10> rsfs7 <10->rsfs10}{} 
\DeclareMathAlphabet{\mycal}{OT1}{rsfs}{m}{n}

\DeclareMathOperator{\extdm}{d}
\newcommand{\extd}{\extdm \!}

\newcommand{\vecX}{\boldsymbol{X}}
\newcommand{\vecJ}{\boldsymbol{J}}

\newcommand{\algorithm}[1]{\paragraph{}{\bf #1}}

\newcommand{\dtdot}[1]{\dot{\ddot{#1}}}

\begin{document}

\hfill {\footnotesize TUW--10--03}

%\hfill {\footnotesize \today}

\title{All stationary axi-symmetric local solutions of topologically massive gravity}

\author{Sabine Ertl, Daniel Grumiller and Niklas Johansson}

\address{Institute for Theoretical Physics,
           Vienna University of Technology,\\
           Wiedner Hauptstr.~8-10/136,
           Vienna, A-1040, Austria} 

\eads{\mailto{sertl@hep.itp.tuwien.ac.at}, \mailto{grumil@hep.itp.tuwien.ac.at}, \mailto{niklasj@hep.itp.tuwien.ac.at}}
 %}

\begin{abstract}
We classify all stationary  axi-symmetric solutions of topologically massive gravity into Einstein, Schr\"{o}dinger, warped and generic solutions. We construct explicitly all local solutions in the first three sectors and present an algorithm for the numerical construction of all local solutions in the generic sector. The only input for this algorithm is the value of one constant of motion if the solution has an analytic centre, and three constants of motion otherwise. We present several examples, including soliton solutions that asymptote to warped AdS.
\end{abstract}


\setcounter{footnote}{0}

\section{Introduction}

Topologically massive gravity \cite{Deser:1982vy,Deser:1982sv} %,Deser:1982wh,Deser:1982a} 
(TMG) is a 3-dimensional theory of gravity that exhibits gravitons 
and, for negative cosmological constant, black holes \cite{Banados:1992wn}. Its action is given by
\eq{
S_{\textrm{\tiny TMG}} = -\int \extd^3x\sqrt{|g|}\,\Big(R+\frac{2}{\ell^2}\Big) - \frac{1}{2\mu}\,\int\extd^3x \epsilon^{\la\mu\nu}\,\Ga^\si{}_{\la\rho}\,\Big(\partial_\mu\Ga^\rho{}_{\nu\si}+\frac23\,\Ga^\rho{}_{\mu\tau}\Ga^\tau{}_{\nu\si}\Big)
}{eq:intro1}
The interest in TMG was rekindled in 2008 due to a paper by Li, Song and Strominger \cite{Li:2008dq}, and the discussion that ensued after the papers by Carlip, Deser, Waldron and Wise \cite{Carlip:2008jk} and two of the current authors \cite{Grumiller:2008qz}. %,Ertl:2009ch}.

In this work we continue the TMG saga by explaining how to construct all its stationary, axi-symmetric solutions. 
Finding non-trivial solutions to the TMG equations of motion is a very difficult problem and thus far few solutions are known.
For a recent summary and algebraic classification of all known exact solutions see \cite{Chow:2009km} %,Chow:2009vt} 
and references therein.
As remarked in \cite{Chow:2009km}, %,Chow:2009vt}, 
one important reason for the scarceness of known solutions is 
the presence of no-go theorems. For instance, there are no non-trivial static solutions. In fact all solutions that have a (non-null) hypersurface-orthogonal Killing vector are Einstein \cite{Aliev:1996eh}. %, Cavaglia:1999si, Deser:2009er}.

A weaker assumption that can aid in finding solutions is to require instead two commuting Killing vectors. Such spacetimes are the focus of this paper. In this case a powerful method pioneered by Clement \cite{Clement:1994sb} becomes available: stationary axi-symmetric TMG can be formulated in terms of a 0+1 dimensional theory. We call this theory ``topologically massive mechanics'' (TMM). Despite its superficial simplicity it is surprisingly difficult to find analytic solutions of this theory. 
On the other hand there are some non-existence results: for the tuning $|\mu\ell| = 1$ all solutions that satisfy Brown-Henneaux boundary conditions are Einstein \cite{Maloney:2009ck} and for the case $|\mu\ell| = 3$ all solutions that satisfy a set of asymptotically warped boundary conditions are the null-warped black holes (there is an anlog statement for $\mu\ell>3$ asymptotic Schr\"odinger black holes) \cite{Anninos:2010pm}.
Let us classify the solutions of TMM
into four sectors, restricting for the time being to negative cosmological constant $\ell^2>0$.
\begin{enumerate}
\item {\bf Einstein:} these solutions solve also the 3-dimensional Einstein equations; the only solutions in this sector are locally AdS$_3$ and obey Brown--Henneaux boundary conditions
\item {\bf Schr\"{o}dinger:} these solutions are either asymptotically Schr\"{o}dinger or asymptotically AdS$_3$. They behave like zero modes of pp-waves. %some asymptote to AdS$_3$ but violate Brown--Henneaux boundary conditions 
\item {\bf Warped:} these solutions are locally or asymptotically warped AdS$_3$
\item {\bf Generic:} all solutions that are neither Einstein, nor Schr\"{o}dinger, nor warped
\end{enumerate}
The three first categories were essentially exhausted in \cite{Clement:1994sb} and correspond to all presently known solutions of TMM.

In this work we analyse TMM and its space of solutions by several methods. First we perform a Hamiltonian analysis and count the dimension of the physical phase space. We find that the phase space in general is six-dimensional. However, we find 
an interesting four-dimensional sub-space of the physical phase space occurring naturally. This subspace is generated by two conditions that are preserved by the equations of motion. Assuming that these conditions hold reduces the dimension of the physical phase space to four. For this special subspace the Hamiltonian dynamics simplify considerably and we are able to construct all local solutions in closed form. In fact, we prove that one is left with exactly the three first sectors: Einstein, Schr\"{o}dinger and warped. The generic sector then contains all solutions outside this sub-space.

To describe the generic sector we resort to a numerical analysis. It turns out that one needs to employ three 
slightly different methods depending on the initial data. We provide a complete algorithm for numerically constructing
a solution given any consistent initial data. 
The numerical solutions that we find in the general sector have non-constant curvature invariants. Some, but not all, of the examples we construct develop naked curvature singularities. In the set of smooth examples we discuss in particular soliton solutions that asymptote to warped AdS and unravel their asymptotic (non-analytic) behaviour. We also generalise many of our results to TMG with vanishing or positive cosmological constant and address possible applications to recent new massive gravity theories.

This work is organised as follows: In Section \ref{sec:2} we review the derivation of TMM and provide a useful reformulation of the equations of motion. In Section \ref{sec:3} we perform a Hamiltonian analysis of TMM and count the dimension of the physical phase space. %and construct its constraints.
In Section \ref{sec:4} we classify the solutions into Einstein, Schr\"{o}dinger, warped and generic and prove that the generic sector is identical to the solutions violating the two conditions found in the Hamiltonian analysis. We provide simple algorithms to construct local solutions for all possible cases in Section \ref{sec:5}. In Section \ref{sec:examples} we give several examples and show in particular that the generic sector is not empty. An outstanding class of examples are soliton solutions that asymptote to warped AdS. In Section \ref{sec:gen} we provide generalisations to the case with vanishing or positive cosmological constant and give directions for future work.

Before starting we list our conventions. We use signature $(+,-,-)$. For the target space of TMM we bring the 3-dimensional Minkowski metric into light-cone gauge $2\eta_{+-}=-\eta_{YY}=1$. We fix the sign of the target space epsilon-tensor by choosing $\epsilon_{+-Y}=+\frac12$. To reduce clutter we drop overall factors in front of actions, since we are only interested in equations of motion and their solutions. Our sign convention for the Ricci-scalar is such that $R<0$ for AdS.
When we write Lorentz vectors as row or column vectors the components of these vectors always have upper indices. We define the cross product between two Lorentz vector as $({\bf A} \times {\bf B})^i = \epsilon^{i}{}_{jk} A^j B^k$.

\section{Topologically massive mechanics}\label{sec:2}

Following Clement \cite{Clement:1994sb} (see also \cite{Moussa:2003fc,Bouchareb:2007yx,Maloney:2009ck}) we employ the following gauge for a generic stationary, axi-symmetric 3-dimensional line-element
\begin{subequations}
\label{eq:clement}
\eq{
\extd s^2=g_{\mu\nu}\,\extd x^\mu\extd x^\nu + \frac{e^2}{\vecX^2}\,\extd\rho^2
}{eq:as1}
with the 2-dimensional metric
\eq{
g_{\mu\nu}=\left(\begin{array}{cc}
X^+ & Y \\
Y & X^-
\end{array} \right)_{\mu\nu}
}{eq:as2}
whose determinant is given by the norm of the Lorentzian 3-vector $\vecX=(X^+,\,X^-,\,Y)$
\eq{
\textrm{det}\,g = \vecX^2 = X^i X_i = X^i X^j \eta_{ij} = X^+X^- - Y^2 
}{eq:as3}
\end{subequations}
All quantities depend on the radial coordinate $\rho$ only. The vector $\vecX$ must be spacelike everywhere, possibly except at boundaries of the range of definition of $\rho$ where it can become null. These boundaries can correspond to curvature singularities, Killing horizons, the radial centre or the asymptotic boundary of spacetime. For brevity, we use the term 
``centre'' to refer to any point where $\vecX^2 = 0$.

Plugging the Ansatz for the line-element \eqref{eq:clement} into the action \eqref{eq:intro1} and integrating out the Killing coordinates $x^\mu$ yields a reduced action in 0+1 dimensions:
\eq{
S_{\textrm{TMM}} =\int\extd\rho\, e\, \Big(\frac12\,e^{-2}\,\dot\vecX^2 - \frac{2}{\ell^2} - \frac{1}{2\mu}\,e^{-3}\,\epsilon_{ijk}\,X^i\dot X^j\ddot X^k\Big)
}{eq:as4}
Dots denote derivatives with respect to the radial coordinate $\rho$, which acts as ``time'' in the particle mechanics system described by the TMM action \eqref{eq:as4}. Any solution to the field equations descending from the action \eqref{eq:as4} can be oxidised to a solution of TMG by virtue of the line-element \eqref{eq:clement}. All stationary axi-symmetric solutions of TMG can be obtained locally in this way \cite{Clement:1994sb}.

For $\mu\to\infty$ the action \eqref{eq:as4} simplifies to the geodesic action. For finite $\mu$ this geodesic action is corrected by a non-linear higher derivative interaction originating from the Chern--Simons term. TMM describes a spacelike particle moving in an auxiliary 3-dimensional Minkowski target space. The Einbein $e$ in the TMM action \eqref{eq:as4} acts as a Lagrange multiplier. We set it to unity after variation, without loss of generality. This amounts to a gauge fixing of the $\rho$-reparameterisation invariance. Variation with respect to the Einbein yields the Hamilton constraint
\eq{
G = \frac12\,\dot\vecX^2 + \frac{2}{\ell^2} - \frac{1}{\mu}\,\epsilon_{ijk}\,X^i\dot X^j\ddot X^k  = 0
}{eq:as5}
Variation with respect to the target space coordinates $X^i$ establishes the equations of motion
\eq{
\ddot X_i = -\frac{1}{2\mu}\,\epsilon_{ijk} \,\big(3\dot X^j \ddot X^k+2X^j\dtdot{X}^k\big)
}{eq:as6}
Invariance of the TMM action \eqref{eq:as4} under Lorentz transformations leads to the conserved angular momentum $\vecJ$, which provides first integrals to the equations of motion \eqref{eq:as6}. 
\eq{
\vecJ =  \vecX \times \dot\vecX + \frac{1}{\mu}\,\vecX\times(\vecX\times\ddot\vecX)-\frac{1}{2\mu}\,\dot\vecX\times(\vecX\times\dot\vecX)
}{eq:as6.5}
We provide now a useful reformulation of the equations of motion. The conserved angular momentum can be represented as
\eq{
J_i = \epsilon_{ijk}\,X^j\dot X^k -\frac{1}{4\mu}\,\big(5\dot\vecX^2+\frac{12}{\ell^2}\big)\,X_i+\frac{1}{2\mu}\,(\vecX\dot\vecX)\,\dot X_i-\frac{1}{\mu}\,\vecX^2\ddot X_i
}{eq:as7}
Combining the Hamilton constraint  \eqref{eq:as5} conveniently with the first integrals \eqref{eq:as7} yields the condition
\eq{
%\vecX\ddot\vecX=\frac{3}{4}\,\big(\dot\vecX^2+\frac{4}{\ell^2}\big)
\epsilon_{ijk}\,J^i X^j\dot X^k = \frac12\,\vecX^2\dot\vecX^2-(\vecX\dot\vecX)^2-\frac{2}{\ell^2}\,\vecX^2
}{eq:as8}
Contracting the angular momentum $\vecJ$ with $\vecX$ and using the equations of motion leads to the constraint
\eq{
\vecX \vecJ = \frac{1}{2\mu}\,\big((\vecX\dot\vecX)^2-\vecX^2\dot\vecX^2\big) 
}{eq:as9}
Given some initial data $\vecX$, $\dot\vecX$ and values for the constants of motion $\vecJ$ compatible with the conditions \eqref{eq:as8} and \eqref{eq:as9} the first integrals \eqref{eq:as7} provide an evolution equation for $\ddot\vecX$. Note that no third [second] derivative terms are present in \eqref{eq:as7} [\eqref{eq:as8}, \eqref{eq:as9}].

The equation for the angular momentum \eqref{eq:as7} together with the constraints \eqref{eq:as8}, \eqref{eq:as9} fully describe the dynamics of the theory defined by the TMM action \eqref{eq:as4} at all points, except at singularities of the line-element \eqref{eq:clement}. Our goal is to find all solutions of these equations, and thus locally all stationary, axi-symmetric solutions to TMG. We start with a counting of the physical degrees of freedom. 

\section{Hamiltonian analysis}\label{sec:3}

To determine the dimension of the physical phase space we perform a constraint analysis in a suitable Hamiltonian formulation. We rewrite the TMM action \eqref{eq:as4} in first order form:
\eq{
S_{\textrm{TMM}} =\int\extd\rho \,\Big(p^x_i\dot x^i+p^y_i\dot y^i - H(e,x^i,y^i,z^i,p^x_i,p^y_i)\Big)
}{eq:as10}
Our phase space is 20-dimensional, with canonical coordinates $(e,x^i,y^i,z^i)$ and canonical momenta $(p^e,p^x_i,p^y_i,p^z_i)$. The quantity $x^i$ is identical to $X^i$, while $y^i$ and $z^i$ for $e=1$ correspond to the velocity $\dot\vecX$ and acceleration $\ddot\vecX$, respectively. Similar to the canonical analysis of TMG at the logarithmic point\footnote{This is also known as ``chiral point'' or ``critical point''.} $|\mu\ell|=1$ \cite{Grumiller:2008pr} we have identified already the appropriate canonical momenta in order to avoid a larger phase space with more constraints, since such an approach eventually would reduce to the formulation \eqref{eq:as10} that we are using as starting point (see also \cite{Faddeev:1988qp}). The momenta $p^x_i$ and $p^y_i$ are in fact Lagrange multipliers enforcing the relations $\dot{x}^i  = e\, y^i$ and $\dot{y}^i=e\, z^i$. The canonical Hamiltonian is given by
\eq{
H = e\, G
}{eq:as11}
with the Hamilton constraint
\begin{subequations}
\label{eq:constraints}
\eq{
G = p^x_i y^i + p^y_i z^i - \frac12\,y_i y^i+\frac{2}{\ell^2}+\frac{1}{2\mu}\,\epsilon_{ijk}\,x^i y^j z^k
}{eq:as12}
Defining the quantities
%\begin{align}
%\Pi_i &:= p^y_i + \frac{1}{2\mu}\,\epsilon_{ijk}\,x^j y^k \\
%\Phi_i &:= p^x_i - y_i - \frac{1}{\mu}\,\epsilon_{ijk}\,x^j z^k \\
%\Psi_i &:= z_i + \frac{3}{2\mu}\,\epsilon_{ijk}\,y^j z^k
%\end{align}
\eq{
\Pi_i := p^y_i + \frac{1}{2\mu}\,\epsilon_{ijk}\,x^j y^k\qquad \Phi_i := p^x_i - y_i - \frac{1}{\mu}\,\epsilon_{ijk}\,x^j z^k \qquad \Psi_i := z_i + \frac{3}{2\mu}\,\epsilon_{ijk}\,y^j z^k
}{eq:angelinajolie}
we obtain the following set of constraints ($\approx$ means weakly vanishing):
\begin{align}
\textrm{Primary:} \quad & p^e \approx 0 &\qquad \textrm{Primary:} &  & p^z_i \approx 0 &\qquad& \textrm{Hamilton:} \quad & G \approx 0 \\ 
\textrm{Secondary:} \quad & \Pi_i \approx 0 &\qquad \textrm{Ternary:} &  & \Phi_i \approx 0 &\qquad& \textrm{Quaternary:} \quad & x^i\Psi_i \approx 0
%\textrm{Primary:} \quad & p^e \approx 0 \\
%\textrm{Primary:} \quad & p^z_i \approx 0 \\
%\textrm{Hamilton:} \quad & G \approx 0 \\
%\textrm{Secondary:} \quad & \Pi_i \approx 0 \\
%\textrm{Ternary:} \quad & \Phi_i \approx 0 \\
%\textrm{Quaternary:} \quad & x^i\Psi_i \approx 0 
\end{align}
\end{subequations}
Two linear combinations of these twelve constraints are first class: $p^e$ and the extended Hamiltonian $H^{\rm ext}$
\eq{
H^{\rm ext} = e\, \big(G + \la^i p^z_i\big)
}{eq:as13}
where the Lagrange multipliers $\la^i$ are determined functions of the canonical variables. All other constraints are second class. We collect in \ref{app:A} the non-vanishing Poisson brackets between the constraints \eqref{eq:app1}, results for the Lagrange multipliers $\la^i$ \eqref{eq:app2} and the Hamilton equations of motion \eqref{eq:Heom}. We conclude that the constraints eliminate fourteen degrees of freedom and we are left with a six-dimensional physical phase space.

The physical phase space has an interesting subspace. Namely, suppose that for some configuration all constraints \eqref{eq:constraints} are fulfilled and in addition the following constraints hold at some value of the ``time'' $\rho$
\eq{
\textrm{Assumptions:}\qquad y^i\Psi_i = y^i z_i \approx 0 \qquad z^i\Psi_i = z^i z_i \approx 0
}{eq:as14}
Then these fourteen constraints hold for all times (see \ref{app:A}). If $x^i, y^i, z^i$ are linearly independent then the extended Hamiltonian \eqref{eq:as13} reduces to the canonical one \eqref{eq:as11}. The subspace of the physical phase space defined by the additional constraints \eqref{eq:as14} is only four-dimensional. Interestingly, all currently known stationary, axi-symmetric solutions to TMG belong to this subspace. 

We note that the angular momentum \eqref{eq:as7} takes a rather simple form in terms of canonical variables.
\eq{
J_i \approx \epsilon_{ijk}\,\big(x^j p^{x\,k}+y^j p^{y\,k}\big)
}{eq:as15}
The angular momentum commutes with the canonical Hamiltonian \eqref{eq:as11}, with the extended Hamiltonian \eqref{eq:as13}, with the primary constraints $p^e, p^z_i$ and with the secondary constraints $\Pi_i$. The constraints \eqref{eq:constraints} and the Hamilton equations of motion \eqref{eq:Heom} taken together are equivalent to the Lagrangian equations of motion \eqref{eq:as7}-\eqref{eq:as9}. 

In the next section we shall freely switch between the Hamiltonian and the Lagrangian formulation, depending on which description is most adequate to illuminate certain points. To facilitate this switching we always employ the gauge $e=1$.
\eq{
\textrm{Gauge\;fixing condition:}\quad e=1
}{eq:as16}
In this gauge $X^i=x^i$, $\dot X^i=y^i$ and $\ddot X^i=z^i$.

\section{Classification of all stationary axi-symmetric solutions}\label{sec:4}

We consider here all solutions to the equations of motion \eqref{eq:as5}-\eqref{eq:as6} [or, equivalently, of \eqref{eq:as7}-\eqref{eq:as9}]. We classify the solutions into four sectors: Einstein, Schr\"{o}dinger, warped and generic.

To simplify certain expressions we shall often exploit Lorentz transformations, constant rescalings
\eq{
\rho\to \Omega \,\rho\qquad \vecX\to\Omega^{-1}\vecX\qquad\dot\vecX\to\dot\vecX\qquad\ddot\vecX\to\Omega\,\ddot\vecX
}{eq:as31}
and constant shifts of $\rho$, all of which leave invariant the line-element \eqref{eq:clement} up to diffeomorphisms (and possibly up to changes of ranges of definition of spacetime coordinates). The rescalings \eqref{eq:as31} together with the $SL(2,\mathbb{R})$ transformations corresponding to Lorentz boosts and rotations form the group $GL(2,\mathbb{R})$.

\subsection{Einstein sector}

This sector encompasses all solutions with vanishing acceleration, $\ddot\vecX=0$.

The geodesic equations of motion are supplemented by a simple constraint on the norm of the velocity
\eq{
\ddot\vecX = 0 \qquad \dot\vecX^2 = - \frac{4}{\ell^2}
}{eq:as29}
The solution to the equations of motion \eqref{eq:as29} is given by
\eq{
\vecX = \vecX_{(0)} \,\rho + \vecX_{(2)} 
}{eq:as22}
with some constant vectors $\vecX_{(0)}, \vecX_{(2)}$, where $\vecX_{(0)}^2=-\frac{4}{\ell^2}$. These solutions constitute all stationary, axi-symmetric solutions of Einstein gravity, as pointed out first by Clement \cite{Clement:1992ke}. By exploiting Lorentz transformations, rescalings of $\rho$ and shifts of $\rho$ we can bring $\vecX$ into one of the following forms ($a\neq 0$)
\eq{
\vecX = \big(a,\;\frac{1}{a},\;\pm\frac2\ell\,\rho + 1\big) \qquad  \vecX = \big(a,\;0,\;\pm\frac2\ell\,\rho\big) %\qquad \vecX = \big(0,\;a,\;\pm\frac2\ell\,\rho\big)  
}{eq:as33}
The corresponding conserved angular momentum is given by
\eq{
\vecJ = \frac{2a}{\mu\ell^2}\;\big( 1\mp\mu\ell,\;\frac{1}{a^2}(1\pm\mu\ell),\;0\big) \qquad \vecJ = \frac{2a}{\mu\ell^2}\;\big( 1\mp\mu\ell,\;0,\;0\big) %\qquad \vecJ = \frac{2a}{\mu\ell^2}\;\big(0,\;1\pm\mu\ell,\;0\big) \qquad 
}{eq:as37}
At $\rho=0$ the coordinate vector \eqref{eq:as33} becomes light-like. %This locus corresponds to the radial centre of spacetime. 
Note that by boosting in the $(X^+X^-)$-plane it is possible to eliminate also the single remaining constant of motion $a$. Thus all solutions are locally equivalent. We keep $a$ arbitrary since it makes it simpler to connect \eqref{eq:as33} with the BTZ metric. The angular momentum \eqref{eq:as37} becomes light-like at the logarithmic point $|\mu\ell|=1$ for the first case and is always light-like for the second case; we shall see below that the latter corresponds to extremal BTZ black holes.

We discuss now briefly how to obtain the BTZ metric. Inserting the left solution \eqref{eq:as33} into the line-element \eqref{eq:clement} one obtains
\eq{
\extd s^2= a\,(\extd x^+)^2 + \frac1a\,(\extd x^-)^2 \pm \big(e^{2r} \frac1\al + e^{-2r}\al\big)\,\extd x^+\extd x^- - \ell^2\extd r^2 
}{eq:as67}
where $\rho=\ell\, (e^{2r}/\al+e^{-2r}\al\mp 2)/4$ with $\al=4G(L\bar L)^{1/2}/\ell$. The arbitrary constant $\al$ parameterises the angular coordinate $\phi=(x^+/\al\ell^2\pm x^-)/2$, which has periodicity of $2\pi$. Defining new coordinates by $x^-=\mp v$, $x^+=\ell^2\al\, u$ and parameterising the constant of motion as $1/a=4G\ell\bar L$ yields
\eq{
\extd s^2= -4G\ell\big(L\extd u^2+\bar L\extd v^2\big)-\big(\ell^2e^{2r}+16G^2L\bar L e^{-2r}\big)\extd u \extd v -\ell^2\extd r^2 
}{eq:as68}
This is one of the standard ways to represent the BTZ black hole metric \cite{Banados:1998gg}, %,Carlip:2005zn}, 
where $L, \bar L$ are related to outer/inner horizon $r_\pm$ and mass/angular momentum $m, j$ as follows:
\eq{
L=\frac{(r_+ + r_-)^2}{16G\ell}\qquad \bar L=\frac{(r_+ - r_-)^2}{16G\ell}\qquad\qquad m=L+\bar L\qquad j=L-\bar L
}{eq:as69}
The extremal case $\bar L=0$ arises as a singular limit of the first solution \eqref{eq:as33} and is described adequately by the second solution \eqref{eq:as33}. Global (Poincar{\'e} patch) AdS arises in the limit $j\to 0$, $m\to-1/8G$ ($j\to 0$, $m\to 0$). 
The results above are obviously in accordance with the local triviality of 3-dimensional Einstein gravity \cite{Deser:1984tn}.
All solutions in the Einstein sector are locally and asymptotically AdS. 

We shall encounter solutions in the other sectors that are also asymptotically AdS. In TMG there are several possible behaviours for the subleading terms. A general Fefferman--Graham like expansion for the coordinate vector in the limit of large $\rho$ takes the form
\eq{
\vecX = \vecX_{(0)}\,\rho + o(\rho)
}{eq:as30}
In TMG any vector $\vecX$ compatible with the expansion \eqref{eq:as30} leads to a 3-dimensional spacetime that is asymptotically AdS. If the subleading terms $o(\rho)$ are constant then the more restrictive Brown--Henneaux boundary conditions are fulfilled. (This eliminates certain asymptotically AdS solutions of TMG.) These boundary conditions hold for \eqref{eq:as22}, in accordance with \cite{Brown:1986nw}. If $o(\rho)$ is logarithmic in $\rho$ then the logarithmic boundary conditions of \cite{Grumiller:2008es,Henneaux:2009pw} are required. We shall encounter such a solution below. If $o(\rho)$ behaves like $\rho^{1-\eps}$, with $\eps>0$, then yet other boundary conditions apply \cite{Henneaux:2009pw}. We give examples of such behaviour when describing the Schr\"{o}dinger sector below.

\subsection{Schr\"{o}dinger sector}\label{sec:L}

This sector encompasses all solutions where the coordinate vector $\vecX$, its velocity $\dot\vecX$ and its acceleration $\ddot\vecX$ are linearly dependent and acceleration is non-vanishing, $\ddot\vecX\neq 0$.

If $\vecX$ and $\dot\vecX$ are parallel then the equations of motion have no solution with $\ddot\vecX \neq 0$, so we can disregard this case. Thus, in order to get linear dependence between coordinate vector, velocity and acceleration we assume with no loss of generality
\eq{
\ddot\vecX = A(\rho)\,\vecX + B(\rho)\,\dot\vecX
}{eq:as17}
Then all triple products between any triple of coordinate vectors or its time derivatives vanish, and we obtain from the Hamiltonian constraint and its time derivative the conditions
\eq{
\dot\vecX^2 = - \frac{4}{\ell^2}\qquad \dot\vecX\ddot\vecX = 0
}{eq:as18}
Contracting the equations of motion \eqref{eq:as6} with the coordinate vector and its acceleration establishes
\eq{
\vecX\ddot\vecX = 0 \qquad \ddot\vecX^2 = 0
}{eq:as19}
The second equations in \eqref{eq:as18}, \eqref{eq:as19} imply that the assumptions \eqref{eq:as14} are fulfilled. Contracting the angular momentum \eqref{eq:as7} with the acceleration yields
\eq{
\ddot\vecX\vecJ = 0
}{eq:as20}
Since the angular momentum is constant the last condition implies that the acceleration must be proportional to some overall function times a constant vector that has vanishing inner product with the angular momentum. We exploit Lorentz transformations to bring the light-like acceleration into a simple form.
\eq{
\ddot\vecX = \ddot C(\rho)\,\big(1,\;0,\;0\big)
}{eq:as21}
If $\ddot C$ vanishes we are back to the Einstein sector discussed above.
If $\ddot C$ is non-vanishing we can integrate the acceleration \eqref{eq:as21} and impose the Hamiltonian constraint to determine $\dot\vecX$:
\eq{
\dot\vecX = \big(\dot C(\rho),\; 0,\; \pm\frac{2}{\ell}\big)
}{eq:as23}
Now we integrate the velocity \eqref{eq:as23}, impose the remaining constraints derived above and exploit constant shifts in $\rho$ so that the norm of $\vecX$ is non-vanishing for positive $\rho$:
\eq{
\vecX = \big(C(\rho),\; 0,\; \pm\frac{2}{\ell}\, \rho \big)
}{eq:as24}
Plugging this vector into the equations of motion \eqref{eq:as6} yields a third order ordinary differential equation for $C(\rho)$ that can be solved straightforwardly. Exploiting rescalings of $\rho$ we obtain the final result ($|\mu\ell|\neq 1$)
\eq{
\vecX = \big(s\rho^{(1\mp\mu\ell)/2} + a\rho + b,\; 0,\; \pm\frac{2}{\ell}\, \rho \big) 
}{eq:as25}
which depends on two integration constants $a,b$ and a sign $s=\pm$. Again, $a$ can be fixed to an arbitrary number by further Lorentz transformations and $b$ can be rescaled by a positive number. We keep them arbitrary for convenience.
The result \eqref{eq:as25} leads to 3-dimensional spacetimes with asymptotic Schr\"{o}dinger behaviour:\footnote{The Schr\"{o}dinger solutions were discussed already by Clement \cite{Clement:1994sb}, except for the two logarithmic solutions \eqref{eq:as26}, \eqref{eq:as27} below that were discovered in \cite{AyonBeato:2004fq, AyonBeato:2005qq}. 
Higher-dimensional versions of the asymptotic Schr\"{o}dinger line-element \eqref{eq:as89} have been proposed as gravity duals for cold atoms \cite{Son:2008ye}. %,Balasubramanian:2008dm} 
(See also \cite{Kachru:2008yh} for the holographic description of other condensed matter systems with anisotropic scaling.)} %, Danielsson:2009gi}.} 
\eq{
\extd s^2\big|_{r\to 0} \sim \ell^2\,\Big(\frac{\pm 2\extd x^+\extd x^- -\extd r^2}{r^2}+\beta\, \frac{(\extd x^+)^2}{r^{2z}}\Big)
}{eq:as89}
where $\beta$ is a constant whose explicit value will be given below.
The scaling exponent $z$ is related to the parameters in the action as follows.
\eq{
z = \frac{1\mp\mu\ell}{2}
}{eq:as32} 
For $\mu\ell=\mp 3$ we obtain the special case of (null-)warped AdS, $z=2$. Despite the name ``null warped'' this solution belongs to the Schr\"{o}dinger sector according to our classification since $\vecX$, $\dot{\vecX}$ and $\ddot{\vecX}$ are linearly dependent. The constraint \eqref{eq:as9} implies a light-like angular momentum
\eq{
\vecJ = \big(\frac{2b}{\mu\ell^2}(1\mp\mu\ell),\;0,\;0 \big)
}{eq:as28}
The angular momentum provides one constant of motion $b$.  The periodicity of the angular coordinate $\phi$ constitutes a second free parameter. Thus, again all geometries in this sector are parameterised by two constants, mass and angular momentum. 

Let us now obtain the Schr\"{o}dinger line-element in standard form. We take the solution for $\vecX$ \eqref{eq:as25}, plug it into the definition of the line-element \eqref{eq:clement} and obtain
\eq{
 ds^2=\big(s\rho^{(1\mp\mu\ell)/2}+a\rho+b\big)(\extd x^+)^2 \pm \frac{4\rho}{\ell}\,\extd x^+ \extd x^- - \frac{\ell^2 \extd\rho^2}{4\rho^2}
}{eq:as70}
The linear coordinate transformation $x^+=\al(t\pm\ell\,\phi)$, $x^-=(\pm t-\ell\,\phi)/(2\ell\al)$ together with the choice $a=0$ and the definition $m_\mp:=-2\al^2(s\rho^{(1\mp\mu\ell)/2}+b)$ leads to
\eq{
 ds^2=\left(\frac{2\rho}{\ell^2}-\frac{m_{\mp}}{2}\right)\extd t^2\mp \ell m_{\mp}\extd t\extd\phi -\left(2\rho+\ell^2\,\frac{m_{\mp}}{2}\right)\extd\phi^2-\frac{\ell^2\extd\rho^2}{4\rho^2}
}{eq:as72}
This is precisely the line-element (21) of \cite{Clement:1994sb}. The parameters $c,M$ in that work are related to the constants above by $c=s/b$, $M=-2\al^2b$. The arbitrary constant $\al$ parameterises the angular coordinate $\phi$, which has periodicity $2\pi$.  If the scaling exponent \eqref{eq:as32} is smaller than one, $z<1$, then the line-element \eqref{eq:as70} asymptotes to AdS. If the scaling exponent \eqref{eq:as32} is larger than one, $z>1$, then the asymptotic line-element in the limit $\rho=\ell^3/(2r^2)\to\infty$ takes the simple form \eqref{eq:as89} provided we choose $\beta=s(\ell^3/2)^z/\ell^2$.
If the scaling exponent \eqref{eq:as32} is equal to one, $z=1$, then we are at the logarithmic point $|\mu\ell|=1$ that we have neglected so far.

At the logarithmic point $|\mu\ell|=1$ we obtain two solutions. The first one is given by
\eq{
\vecX = \big(\pm \ln{\rho} + a\rho + b,\; 0,\; 2\mu \rho \big)
}{eq:as26}
The angular momentum has again only a non-vanishing $+$-component given by $J^+=\mp 4\mu$.
The result \eqref{eq:as26} leads to a spacetime that is asymptotically AdS reminiscent of the logarithmic mode discovered in \cite{Grumiller:2008qz}.
The other solution at the logarithmic point is given by
\eq{
\vecX = \big(\pm\rho\ln{\rho} + a\rho + b,\; 0,\; -2\mu \rho \big)
}{eq:as27}
The angular momentum has again only a non-vanishing $+$-component given by $J^+=4\mu b$, consistent with the result \eqref{eq:as28}. The result \eqref{eq:as27} leads to a spacetime that marginally violates the asymptotically AdS conditions \eqref{eq:as30}. It is reminiscent of the sources for operators dual to the logarithmic mode \cite{Skenderis:2009nt,Grumiller:2009mw} within the logarithmic conformal field theory dual to TMG at the logarithmic point.  
Like for $|\mu \ell| \neq 1$ we can fix both $a$ and $b$ in \eqref{eq:as26} and \eqref{eq:as27} by a combination of Lorentz transformations and rescalings \eqref{eq:as31} that leaves invariant the acceleration \eqref{eq:as21}.

\subsection{Warped sector}

This sector encompasses all solutions which obey the constraints \eqref{eq:as14} and where coordinate vector $\vecX$, velocity $\dot\vecX$ and acceleration $\ddot\vecX$ are linearly independent.

Imposing the constraints \eqref{eq:as14} implies
\begin{eqnarray}
\ddot\vecX^2=\dot\vecX\ddot\vecX=\epsilon_{ijk}\,X^i\dot X^j\dtdot{X}^k=\epsilon_{ijk}\,X^i\ddot X^j\dtdot{X}^k=0
\label{eq:as34} \\
\dot\vecX^2=\alpha\qquad\!\frac1\mu\,\epsilon_{ijk}\,X^i\dot X^j\ddot X^k=\frac{2}{\ell^2}+\frac{\al}{2}
\end{eqnarray}
with some constant $\al\neq -4/\ell^2$. Due to our assumption of linear independence we can write
\eq{
\dtdot{\vecX} = A(\rho)\,\vecX+B(\rho)\,\dot\vecX+C(\rho)\,\ddot\vecX
}{eq:as35}
The equalities \eqref{eq:as34} imply $B=C=0$. The $\rho$-derivative of a suitable combination of the equations of motion establishes the constraint
\eq{
\vecX\dtdot{\vecX} = -\frac52\,\dot\vecX\ddot\vecX
}{eq:as36}
that must be valid for all solutions. Therefore, also $A$ must vanish in \eqref{eq:as35} and we obtain the condition
\eq{
\dtdot{\vecX} = 0
}{eq:as38}
The general solution in this sector then is given by
\eq{
\vecX = \vecX_{(-2)}\,\rho^2 + \vecX_{(0)}\,\rho + \vecX_{(2)}
}{eq:as39}
where $\vecX_{(-2)}$ is light-like and orthogonal to $\vecX_{(0)}$; the equations of motion \eqref{eq:as5}, \eqref{eq:as6} put further restrictions on the three constant vectors appearing in the solution \eqref{eq:as39}. The Fefferman--Graham like expansion \eqref{eq:as39} clearly is incompatible with asymptotic AdS behaviour \eqref{eq:as30}. In fact, the solutions \eqref{eq:as39} correspond to warped (squashed or stretched) AdS geometries. Solutions of this type were discovered first by Nutku \cite{Nutku:1993eb}.

We exploit now shifts, Lorentz transformations and rescalings and obtain
\eq{
\vecX = \big(\rho^2-2\rho,\;\frac13\,(\mu^2-\frac{9}{\ell^2}),\;-\frac{2\mu}{3}\,\rho+a\big)
}{eq:as40}
As in the previous cases the constant of motion $a$ can be eliminated by a diffeomorphism together with a change of range of definition of the coordinates. The conserved angular momentum is given by
\eq{
\vecJ=\frac{1}{3\mu\ell^2}\,\big(\mu^2\ell^2-9-8\mu\ell^2a+6\ell^2a^2,\;-2\mu^2-\frac{\mu^4\ell^2}{9}+\frac{27}{\ell^2},\;6\mu-\frac23\,\mu^3\ell^2-9a+\mu^2\ell^2a\big)
}{eq:as50}
The angular momentum becomes light-like for $\mu\ell=\pm 3$ or $\mu=2a\pm\sqrt{a^2+9/l^2}$. In the former case we obtain null warped solutions, in the latter case extremal warped black holes (see below). We note that the null warped solutions actually belong to the Schr\"{o}dinger sector since $\vecX$, $\dot\vecX$ and $\ddot\vecX$ depend linearly on each other.

To obtain the warped line-element, we take the solution for $\vecX$ \eqref{eq:as40} and plug it into the definition of the line-element \eqref{eq:clement},
\eq{
        ds^2=\big(\rho^2-2\rho\big)(\extd x^+)^2+\big(\frac{\mu^2}{3}-\frac{3}{\ell^2}\big)(\extd x^-)^2-\big(\frac{4\mu\rho}{3}-\frac{J}{\mu}\big)\extd x^+\extd x^- -\frac{\extd\rho^2}{\frac{\mu^2+27/\ell^2}{9}\rho^2+\frac M3 \rho+\frac{J^2}{4\mu^2}}
}{eq:as71}
with $M:=2\mu^2-18/\ell^2-4\mu a$ and $J:=2\mu a$.
The coordinate transformation $x^+=t/\alpha$, $x^-=\alpha\phi$ leads to Nutku's solution \cite{Nutku:1993eb}, provided we choose $\alpha^2=\frac{1}{2a}\rho^2-\frac{6\rho}{(\mu^2-9\Lambda)}$ and define $\La:=1/\ell^2$, $r^2:=2\rho$. Other coordinate systems useful for the representation of warped AdS and warped AdS black holes are collected in \cite{Anninos:2008fx}. The two zeros in the denominator of the $\extd\rho^2$ term in the line-element \eqref{eq:as71} correspond to Killing horizons. An extremal Killing horizon emerges if the coupling constants and constant of motion are tuned as $\mu=2a\pm\sqrt{a^2+9/l^2}$. This is precisely one of the conditions for $\vecJ$ \eqref{eq:as50} becoming light-like that we found above.

Warped AdS spacetimes have attracted a lot of interest recently as candidates for stable TMG backgrounds \cite{Anninos:2008fx}.
They arise also as string theory solutions \cite{Detournay:2005fz} %Orlando:2010ay} 
and as solutions to the Einstein-perfect fluid field equations \cite{Gurses:2008wu}.

\subsection{Generic sector}

In the Hamiltonian analysis we encountered the interesting feature that the physical phase space is six-dimensional in general, but only four-dimensional if the constraints $\ddot\vecX^2=\dot\vecX\ddot\vecX=0$ hold [see \eqref{eq:as14}]. These constraints are fulfilled for all sectors discussed so far. Note also that these sectors exhaust all solutions for which \eqref{eq:as14} holds.

The generic sector consists of all solutions violating these constraints. 
\eq{
\textrm{Generic:}\quad \ddot\vecX^2 \neq 0\;\textrm{and/or}\;\dot\vecX\ddot\vecX\neq 0 
}{eq:generic}
It is not clear a priori if there are any solutions in the generic sector. Neither is it clear that this sector is empty.

It is relatively straightforward to prove that no generic solutions exist for $\vecX$ that are finite polynomials in $\rho$.
Indeed, solutions for $\vecX$ that are linear polynomials in $\rho$ lead to the Einstein sector, quadratic polynomials to the warped or Schr\"{o}dinger sector and certain higher order polynomials to the Schr\"{o}dinger sector. We demonstrate in \ref{app:proof} that {\em all} higher order polynomial solutions belong to the Schr\"{o}dinger sector.
Thus, any generic solution must be non-polynomial in $\rho$, which makes it hard to obtain solutions by trial and error.  In the next section we resolve this issue by providing numerical algorithms that allow to construct all local solutions in the generic sector.

%It is relatively straightforward to prove that no generic solutions exist for $\vecX$ that are finite polynomials in $\rho$. Indeed, solutions for %$\vecX$ that are linear polynomials in $\rho$ lead to the Einstein sector, quadratic polynomials to the warped or Lifshitz sector and higher order %polynomials to the Lifshitz sector. Thus, any generic solution must be non-polynomial in $\rho$, which makes it hard to obtain solutions by trial %and error.  In the next section we resolve this issue by providing algorithms that allow to construct all local solutions in the generic sector.

\section{Construction of all stationary axi-symmetric solutions}\label{sec:5}

In the previous section we classified all solutions into Einstein, Schr\"{o}dinger, warped and generic. Moreover, we constructed explicitly all local solutions in the first three sectors. However, we did not discuss the generic sector in detail. The purpose of this section is to fill this gap. Namely, we provide suitable initial data and algorithms to construct numerically all local solutions in the generic sector. We postpone an execution of these algorithms to section \ref{sec:examples} below, where several examples will be presented. %In \ref{app:D} we collect some interesting features of solutions for special values of the dimensionless coupling constant $|\mu\ell|$.

 In the following we exploit shifts, Lorentz transformations and rescalings to simplify the construction of solutions as much as possible. We consider first solutions with an analytic centre at some point, fixed to $\rho=0$. The solutions are parameterised by one constant of motion. Then we study solutions that do not necessarily have an analytic centre. We find that they are parameterised by three arbitrary constants of motion. In this case it is not possible to exploit shifts in $\rho$. Therefore, the solutions with analytic centre correspond to a two-dimensional subset of the three initial data. We provide simple algorithms to construct all local solutions and collect them in \ref{app:alg}.

\subsection{Solutions with analytic centre}\label{sec:generic}

We assume for the time being that there exists a centre, in the sense that $\vecX^2=0$ for some finite $\rho$. %This ``centre'' can either be a true centre, like $\rho=0$ in cylindrical coordinates; or it can correspond to a Killing horizon; or it can be the locus of a curvature singularity (this is not possible for an analytic centre); or it can correspond to an asymptotic boundary. Regardless of the true geometric nature of the condition $\vecX^2=0$ we always refer to it as ``centre'' in the following.
Moreover, we assume that the vector $\vecX$ is analytic in $\rho$ at the centre. We exploit shifts to make $\vecX^2=0$ at $\rho=0$, and Lorentz transformations to bring $\vecX$ into a convenient form. 
\eq{
\vecX\big|_{\rho=0} = \big(1,\;0,\;0\big)
}{eq:as42}
Residual combinations of Lorentz transformations and rescalings can be used to simplify $\dot\vecX$ without changing \eqref{eq:as42}: we can boost in the $X^+X^-$-plane and simultaneously rescale, or we can boost in $Y$-direction, make a spatial rotation and then rescale.

\paragraph{}
We study first the case where $\vecX\dot\vecX \neq 0$ at $\rho=0$. 
In this case our Ansatz for the initial data consistent with the equations of motion \eqref{eq:as7}-\eqref{eq:as9} is
\eq{
\vecX\big|_{\rho=0} = \left(\begin{array}{c}
1 \\ 0 \\ 0
\end{array}
\right) \qquad \dot\vecX\big|_{\rho=0} = \mu\,\left(\begin{array}{c}
a \\ \pm 1 \\ 2
\end{array}
\right) \qquad \vecJ = \mu\,\left(\begin{array}{c}
3-\frac{3}{\mu^2\ell^2}\mp a \\ \frac{1}{4} \\ 0
\end{array} \right)
}{eq:as48}
The evolution equations \eqref{eq:as7} do not immediately determine $\ddot\vecX$, since the last term in those equations is multiplied by $\vecX^2$, which vanishes by our assumptions above at $\rho=0$. However, taking the $\rho$-derivative of the evolution equations and using the on-shell identity 
\eq{
R=2\vecX\ddot\vecX + \frac32\,\dot\vecX^2 = -\frac{6}{\ell^2}
}{eq:as44}
we obtain
\eq{
\epsilon_{ijk}\,X^j\ddot X^k-\frac{9}{8\mu}\,\big(\dot X^2+\frac{4}{\ell^2}\big)\,\dot X_i - \frac{5}{2\mu}\,(\dot\vecX\ddot\vecX)\,X_i - \frac{3}{2\mu}\,(\vecX\dot\vecX)\,\ddot X_i = \frac1\mu\,\vecX^2\,\dtdot{X}_i 
}{eq:as43}
At $\rho=0$ the right hand side of \eqref{eq:as43} vanishes and setting the left hand side to zero determines $\ddot\vecX$ at $\rho=0$ by means of the matrix equation
\eq{
A_{ik} \ddot X^k = \frac{9}{8\mu}\,\big(\dot\vecX^2+\frac{4}{\ell^2}\big)\,\dot X_i \qquad \textrm{at\;}\rho=0
}{eq:as45}
with the matrix
\eq{
A_{ik}:=\epsilon_{ijk}\,X^j-\frac{5}{2\mu}\,X_i\dot X_k-\frac{3}{2\mu}\,(\vecX\dot\vecX)\,\eta_{ik}
}{eq:as46}
The matrix in the equations of motion \eqref{eq:as45}-\eqref{eq:as46} is invertible for $\vecX \dot{\vecX}\neq 0$. Solving these equations establishes $\ddot\vecX$ and $\dtdot{\vecX}$. Taking higher $\rho$-derivatives of \eqref{eq:as43} establishes linear equations for higher derivatives of $\vecX$ at $\rho=0$. We note that the norm of the vector $\ddot\vecX$ calculated in this way does not vanish in general at $\rho=0$. It vanishes only if one of the following conditions holds
\eq{
a = \pm 4\, \big(1-\frac{1}{\mu^2\ell^2}\big)\qquad\textrm{or}\qquad a = \pm\frac{32}{9}
}{eq:as51}
For the same values of $a$ it turns out that the constraint $\dot\vecX\ddot\vecX=0$ holds at $\rho=0$. For $a=\pm 32/9$ we obtain warped solutions, while for $a=\pm 4\ (1-1/\mu^2\ell^2)$ we obtain Einstein solutions.

	The generic sector contains only those solutions where $a$ is not given by the values listed in \eqref{eq:as51}. We present  in \ref{app:alg} an algorithm to construct these solutions (see Algorithm 1). This algorithm involves the inversion of $3\times 3$ matrices as the most complicated step, and therefore is not expensive in terms of computer time.

\paragraph{}
Let us now assume that $\vecX\dot\vecX =  0$ at $\rho=0$.  
Then our Ansatz for the initial data consistent with the equations of motion \eqref{eq:as7}-\eqref{eq:as9}, with no loss of generality, is
\eq{
\vecX\big|_{\rho=0} = \left(\begin{array}{c}
1 \\ 0 \\ 0
\end{array}
\right) \qquad \dot\vecX\big|_{\rho=0} = \mu\,\left(\begin{array}{c}
s \\ 0 \\ a 
\end{array}
\right) \qquad \vecJ = \mu\,\left(\begin{array}{c}
-a + \frac54 a^2-\frac{3}{\ell^2\mu^2} \\ 0 \\ 0
\end{array} \right)
}{eq:as41}
Here $s=\pm1,0$ and $a$ is a constant of motion. In the present case the matrix \eqref{eq:as46} is non-invertible.
We then take the $\rho$-derivative of equation \eqref{eq:as43} and obtain in this way an equation for $\dtdot{\vecX}$ at $\rho=0$:
\eq{
\big(A_{ik}-\frac2\mu\,\vecX\dot\vecX\,\eta_{ik}\big)\, \dtdot{X}^k = \frac{9}{8\mu}\,\big(\dot\vecX^2+\frac{4}{\ell^2}\big)\,\ddot X_i + \frac{9}{4\mu}\,\dot\vecX\ddot\vecX\,\dot X_i  -\dot A_{ik} \ddot X^k \qquad \textrm{at\;}\rho=0
}{eq:as47} 
Solving the equations above with the Ansatz \eqref{eq:as41} turns out to be possible only if the initial data are further constrained. Either $a=\pm 2/\mu\ell$ and $\mu\ell\neq\pm 5$, in which case we find $\ddot\vecX=0$ and recover solutions of the Einstein sector with light-like angular momentum; or $a=2/3$, in which case we find $\dot\vecX\ddot\vecX=\ddot\vecX^2=0$ and recover solutions of the warped sector with light-like angular momentum. Thus, the Ansatz \eqref{eq:as41} does not lead to any solutions of the generic sector, except for $a=2/5$ and $|\mu\ell|=5$. 

Generic solutions with $a=2/5$ and $|\mu\ell|=5$ are rather special. At a certain value of $\rho$, which for convenience we shift to $\rho=0$, simultaneously the following quantities vanish: 
\eq{
\vecX^2=\vecX\dot\vecX=\vecX\ddot\vecX=\epsilon_{ijk}X^i\dot X^j\ddot X^k=0 \qquad \textrm{at\;}\rho=0
}{eq:as83} 
However, $\vecX\dtdot{\vecX}\neq 0$. Conversely, one can prove that all generic solutions compatible with these conditions require $a=2/5$ and $|\mu\ell|=5$. 
These solutions need the specification of a single constant of motion, for instance the value of $\dtdot{X}^-$ at $\rho=0$. %We shall discuss examples of such solutions in section \ref{sec:examples} below.
We present in \ref{app:alg} an algorithm to construct these solutions (see Algorithm 2).  As in the previous case this algorithm involves the inversion of $3\times 3$ matrices as the most complicated step.

\subsection{Completely generic solutions}\label{ssec:generic}

If there is no centre then $\vecX^2<0$ everywhere, and we cannot use shifts to make $\vecX$ light-like at $\rho=0$. If the centre is not analytic we cannot use it straightforwardly as starting point for the time evolution. Finally, even if there is an analytic centre we may choose to start our evolution from a point that is not the centre. In either of these cases we have to assume that $\vecX$ is space-like at our starting point of the evolution, which we shift to $\rho=0$. We can still exploit Lorentz transformations to bring $\vecX$ into a convenient form.

A simple counting reveals that we have now three initial data: we need nine data to specify $\vecX|_{\rho=0}$, $\dot\vecX|_{\rho=0}$ and $\vecJ$. However, these data are constrained by \eqref{eq:as8} and \eqref{eq:as9}. Three Lorentz transformations and a rescaling eliminate four additional data, so we end up with $9-2-4=3$ initial data. By comparison, in the case where we evolve from an analytic centre there are three linearly independent constraints on the initial data \eqref{eq:as7}-\eqref{eq:as9}. Moreover, there is the constraint that $\vecX|_{\rho=0}$ is light-like. Again we have three Lorentz transformations and a rescaling. In total we have $9-3-1-4=1$ initial datum. An equivalent way to see this is as follows: there is only one residual $GL(2,\mathbb{R})$ transformation that leaves invariant the space-like Ansatz $\vecX|_{\rho=0}=(0,\,0,\,1)$, which we can use to fix one of the components of $\dot\vecX|_{\rho=0}$. So we have two initial data residing in the velocity, $a,b$, and one datum residing in the angular momentum, $c^+$. However, there are two residual $GL(2,\mathbb{R})$ transformations that leave invariant the light-like Ansatz $\vecX|_{\rho=0}=(1,\,0,\,0)$,  which we can use to fix two of the components of $\dot\vecX|_{\rho=0}$.  So we have one initial datum residing in the velocity, $a$, and no datum residing in the angular momentum, as it is fully determined by \eqref{eq:as7}.

If the inner product between $\vecX$ and $\dot\vecX$ is non-vanishing we can scale it to a certain value in this way. Then our Ansatz for the initial data, assuming $b\neq 0$,  is
\eq{
\vecX\big|_{\rho=0} = \left(\begin{array}{c}
0 \\ 0 \\ 1
\end{array}
\right) \quad \dot\vecX\big|_{\rho=0} = \mu\,\left(\begin{array}{c}
b \\ a \\ \pm 1 
\end{array}
\right) \quad \vecJ = \mu\,\left(\begin{array}{c}
b\mp \frac b2 + c^+ \\
\frac{1}{b}\,\big(\frac{4}{\mu^2\ell^2}-1 \mp \frac12\,ab+a c^+\big) \\
-\frac{1}{2}\, ab
\end{array} \right)
}{eq:as53}
If $\vecX\dot\vecX=0$ at $\rho=0$ our Ansatz for the initial data, assuming $b\neq 0$, is
\eq{
\vecX\big|_{\rho=0} = \left(\begin{array}{c}
0 \\ 0 \\ 1
\end{array}
\right) \qquad \dot\vecX\big|_{\rho=0} = \mu\,\left(\begin{array}{c}
b \\ a \\ 0 
\end{array}
\right) \qquad \vecJ = \mu\,\left(\begin{array}{c}
b + c^+ \\ 
\frac{1}{b}\,\big(\frac{4}{\mu^2\ell^2}+a c^+\big) \\
-\frac{1}{2}\, ab
\end{array} \right)
}{eq:as58}
The initial data are parameterised by three constants $a, b, c^+$ and are consistent with the constraints \eqref{eq:as8}, \eqref{eq:as9}. 

The special case $b=0$ requires separate discussion. We assume first $a\neq 0$. If the inner product between $\vecX$ and $\dot\vecX$ is non-vanishing then our Ansatz for the initial data is
\eq{
\vecX\big|_{\rho=0} = \left(\begin{array}{c}
0 \\ 0 \\ 1
\end{array}
\right) \quad \dot\vecX\big|_{\rho=0} = \mu\,\left(\begin{array}{c}
0 \\ a \\ \pm 1 
\end{array}
\right) \quad \vecJ = \mu\,\left(\begin{array}{c}
\frac1a\,\big(1-\frac{4}{\mu^2\ell^2}\big) \\
-a\mp\frac a2 +c^- \\
0
\end{array} \right)
}{eq:as77}
If $\vecX\dot\vecX=0$ at $\rho=0$ our Ansatz for the initial data is
\eq{
\vecX\big|_{\rho=0} = \left(\begin{array}{c}
0 \\ 0 \\ 1
\end{array}
\right) \qquad \dot\vecX\big|_{\rho=0} = \mu\,\left(\begin{array}{c}
0 \\ a \\ 0 
\end{array}
\right) \qquad \vecJ = \mu\,\left(\begin{array}{c}
-\frac{4}{a\ell^2\mu^2} \\ 
-a+c^- \\
0
\end{array} \right)
}{eq:as78}
The initial data are parameterised by two constants $a, c^-$ and are consistent with the constraints \eqref{eq:as8}, \eqref{eq:as9}. If additionally $a=0$ then there is no generic solution; instead, we recover solutions of the Einstein sector.

We present in \ref{app:alg} an algorithm to construct generic solutions (see Algorithm 3). This algorithm involves solving numerically a coupled set of three second order ordinary differential equations as the most complicated step. 
%and therefore is not expensive in terms of computer time. 
Of course, one has to check whether the solution obtained is really generic or belongs to some of the special cases discussed above. A simple check to rule out Einstein, Schr\"{o}dinger and warped solutions is to calculate $\ddot\vecX^2$ and $\dot\vecX\ddot\vecX$ at some point. If either of these scalar quantities is non-vanishing the solution necessarily is generic. If both of them vanish at one point they must vanish at all points and the solution is not generic. If the solution has no centre then, obviously, it cannot be a solution with analytic centre. If it has a centre then typically the solution will not be analytic there. This is so, because for some random initial data $a,b,c^+$ the subset of data consistent with an analytic centre is a set of measure zero.

\subsection{Summary of generic sector}

Let us now summarise the structure of the generic sector.
After using rescalings and Lorentz transformations, the full space of solutions is parameterised essentially by three constants appearing in the initial data \eqref{eq:as53}-\eqref{eq:as78}. A 2-dimensional surface\footnote{Note that even if there is just one initial datum in \eqref{eq:as48}, this set corresponds to a two dimensional subset of the initial data of \eqref{eq:as53}-\eqref{eq:as78}. The reason is that one can use any value of $\rho$ as starting point. Put differently, in \eqref{eq:as48} we fixed the shift symmetry in $\rho$ which was left unfixed in \eqref{eq:as53}-\eqref{eq:as78}.} in this space of initial data yields the solutions with an analytic centre \eqref{eq:as48}. On this surface there are curves corresponding to non-generic solutions, either Einstein or warped AdS. Besides the solutions on this surface there are two other curves in the space of initial data corresponding to Schr\"{o}dinger solutions. For odd integer values of $\mu\ell$ one of the Schr\"{o}dinger lines lies on the surface corresponding to solutions with analytic centre. To the best of our knowledge no solutions in the generic sector are known so far.\footnote{An interesting method to solve massive gravity field equations in the presence of a Killing vector was presented recently by G\"urses \cite{Gurses:2010sm}. It is not clear, however, how to obtain solutions of the generic sector in this framework.} In section \ref{sec:examples} below we construct numerically the first examples for generic solutions.  In \ref{app:D} we collect some interesting features of solutions for special values of the dimensionless coupling constant $|\mu\ell|$.

\section{Examples}\label{sec:examples} 

We implement now the algorithms described in the previous section and in \ref{app:alg}. %, and we discuss a few examples for illustration.
For the generic case in subsection \ref{ssec:generic} our numeric algorithm essentially consists of letting Mathematica \cite{Mathematica}
solve the coupled system of equations \eqref{eq:as7} with initial data constrained by \eqref{eq:as8} and \eqref{eq:as9}. When we solve around an analytic centre we implement equivalents of Algorithms 1 and 2 described in \ref{app:alg} to obtain a Taylor series around the centre. Then we use this series to provide initial data with $\vecX^2 \neq 0$ close to the centre, and use the generic algorithm from there on.

First we test our algorithms and recover known solutions in section \ref{sec:ex1}. Then we give examples of generic solutions with non-analytic centre in section \ref{sec:ex2}. In section \ref{sec:ex3} we discover soliton solutions with no centre. Finally, in section \ref{sec:ex4} we focus on soliton solutions with analytic centre.

We plot various quantities for different solutions. Typically, we plot the components of $\vecX$, the quadratic curvature invariant
\eq{
K = R_{\mu\nu}R^{\mu\nu} - \frac{3}{8}R^2 = \frac{1}{2}(\vecX \ddot{\vecX})^2 - \frac{1}{2} \vecX^2 \ddot{\vecX}^2 - \frac{1}{4} \dot{\vecX}^2 (\vecX\ddot{\vecX})- \frac{3}{32}(\dot{\vecX}^2)^2
}{eq:ex1}
and as consistency checks the Hamilton constraint $G$ and the components of the equations of motion
\eq{
{\bf EOM} := \ddot{\vecX} + \frac{1}{2\mu}(3\dot{\vecX}\times\ddot{\vecX} + 2\vecX\times\dtdot{\vecX})
}{eq:ex2}
For solutions in the generic sector we also plot $\vecX^2$, $\dot{\vecX}\ddot{\vecX}$, $\ddot{\vecX}^2$ and the components of $\ddot{\vecX}$. Some important plots are inserted in the main text, whereas others are collected in \ref{app:alg} after the references.

\subsection{Recovering Schr\"{o}dinger and warped}\label{sec:ex1}

As a test of our algorithms we first recover numerically several solutions that are known analytically. We present one Schr\"{o}dinger and one warped example. Both are obtained with the generic Algorithm 3. We have also recovered known solutions by making a Taylor expansion around an analytic centre and then starting a generic evolution from a nearby point, but in order to restrict the volume of the paper we do not include such an example in the plots.

The plots on the left in Fig.~\ref{fig:L} represent a numerically computed Schr\"{o}dinger solution. For comparison, on the right the corresponding exact solution is plotted. This example fits in the general expression \eqref{eq:as25} with $\mu = 0.2$, $\ell = 1$, $a=0$, $b=5$ and $s =+1$. We started the evolution from $\rho = 1$ in \eqref{eq:as25}, but this point is shifted to $\rho = 0$
in the plots. This means that the (non-analytic) centre is at $\rho = -1$. As indicated by the plot of ${\bf EOM}$, the numerics
breaks down at this point. This happens even though the spacetime is smooth there. Indeed, $\rho = -1$
does not represent a curvature singularity as is known from the analytic solution. The breakdown of the numerics close to the centre stems from our using \eqref{eq:as7} as evolution equation.

In the warped example presented in Fig.~\ref{fig:W} similar remarks apply. However the centre (also here shifted to $\rho=-1$)
is analytic. This example can be obtained from \eqref{eq:as40} by setting $\mu = 4$, $\ell = 1$ and $a=2$.

In all cases above the numerical results for the quadratic curvature invariant \eqref{eq:ex1} match with high precision the exact expressions. 
\eq{
K_{\textrm{\tiny Schr\"{o}dinger}} = K_{\textrm{\tiny Einstein}} = -\frac{3}{2\ell^4}\qquad\qquad K_{\textrm{\tiny warped}} = \bar K = \frac{4\mu^4\ell^4-72\mu^2\ell^2+243}{54\ell^4}
}{eq:K}
Let us now turn to our real goal in this section: describing solutions that are neither Einstein, nor Schr\"{o}dinger nor warped. 

\subsection{Generic solution with non-analytic centre --- Naked singularities}\label{sec:ex2}

A situation that occurs frequently when one chooses random initial data is that the solution develops curvature singularities.
These typically manifest themselves in the form of non-analytic centres. In Fig.~\ref{fig:2C} we show a representative example of this class of solutions.

For this solution we chose $\mu = 7$, $\ell = 1$, $\vecX(0) = (6,5,7)$, $\dot{\vecX}(0) = (6,7,5)$ and $J^+ = 8$. With these initial data the solution develops two non-analytic centres at $\rho \approx -3.3$ and $\rho \approx 1.21$. That they are centres is clear from the plot of $\vecX^2$. Furthermore, the square invariant $K$ diverges at these points. Thus we have a region of smooth spacetime between two naked curvature singularities. 

As is clear from the plots of $\dot{\vecX}\ddot{\vecX}$ and $\ddot{\vecX}^2$ this solution belongs to the generic sector. Not surprisingly, the numerics is not reliable close to the singular points.

\subsection{Generic solution with no centre --- Solitons}\label{sec:ex3}

\begin{figure}[ht]
\begin{center}
{\bf Soliton with no centre --- Localisation around $\rho\approx -0.5$}\\
\includegraphics[height=180mm]{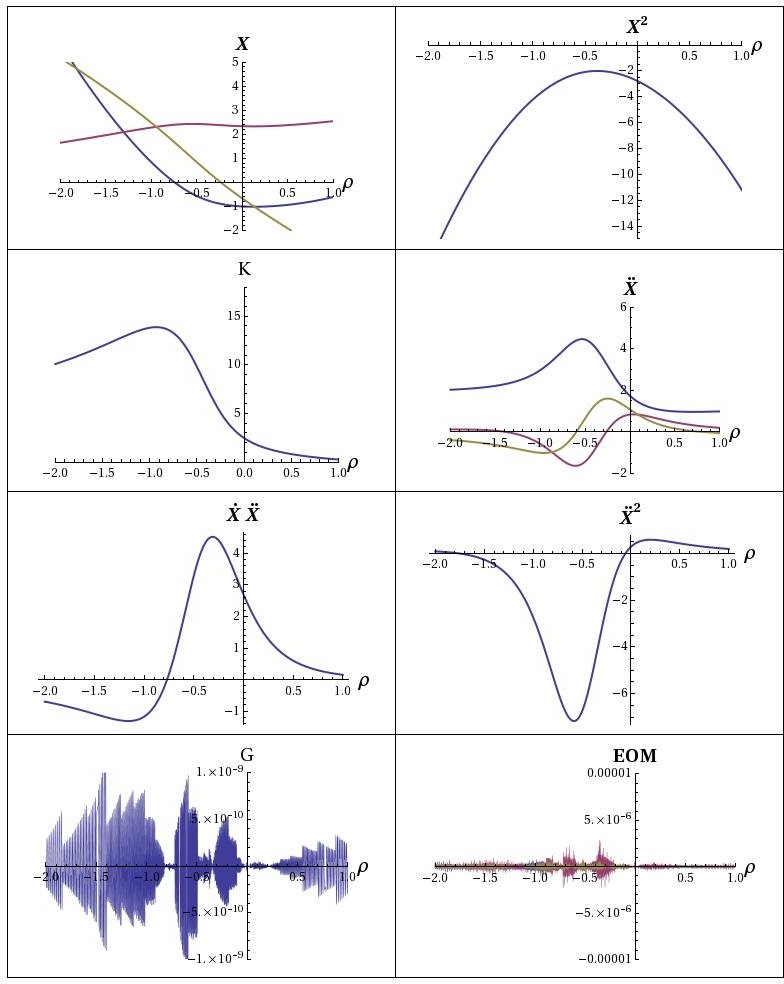}
\caption{The plots show a solution in the generic sector without a centre. Here $\mu = 4$ and $\ell = 1$. Evolution starts at $\rho = 0$ where the initial values $\vecX(0) = (-1, 7/3, -2/3)$, $\dot{\vecX}(0) = (-1/5,-1/10,-8/3)$ and $J^+ = -11/4$ were chosen. \label{fig:Geon1}}
\end{center}
\newpage
\end{figure}
\afterpage{\clearpage}

\begin{figure}[ht]
\begin{center}
{\bf Soliton with no centre --- Evidence for asymptotic warped AdS behaviour}\\
\includegraphics[height=180mm]{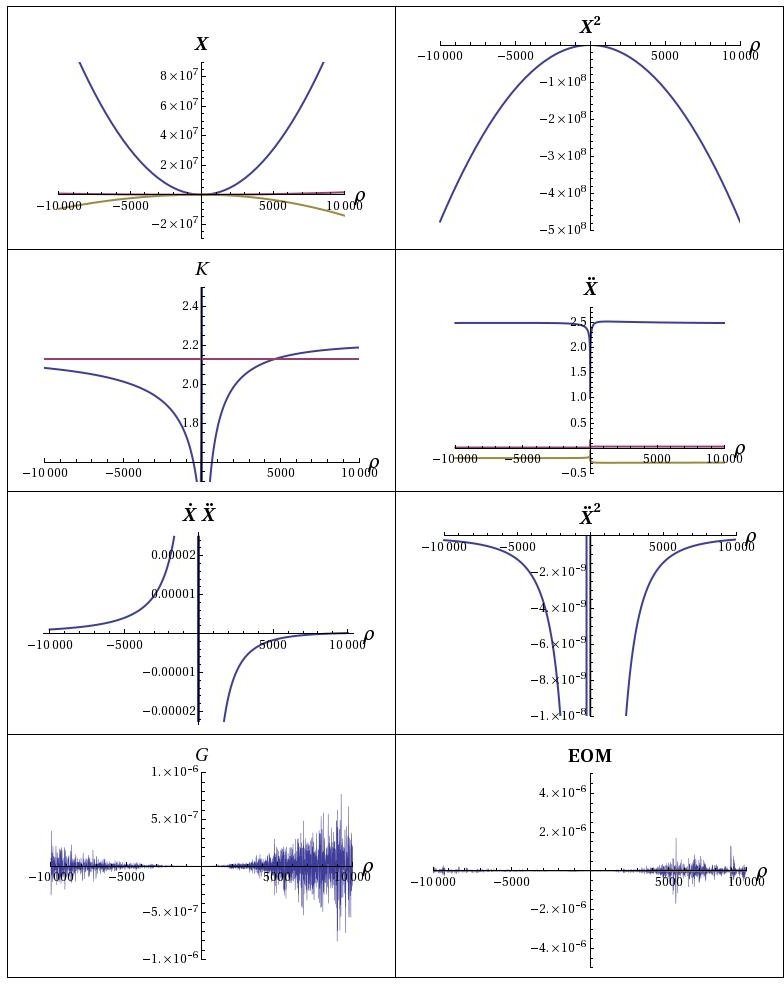}
\caption{The same solution as in Fig.~\ref{fig:Geon1} but with a much larger range in $\rho$. We have let it evolve until the numerics becomes dubious. The facts that $\ddot{\vecX}$ approaches a constant and that $K$ tends to the exact value for warped AdS at $\mu \ell = 4$ (straight line) indicate that the solution is asymptotically warped AdS. \label{fig:Geon2}}
\end{center}
\newpage
\end{figure}

A class of interesting solutions consists of spacetimes without centre, i.e., for which $\vecX^2 < 0$ everywhere.
Varying the initial data by trial and error, it is not hard to obtain such solutions. Typically, there are two asymptotic regions where the constraints \eqref{eq:as14} approach zero, and a central region where they deviate from zero appreciably. 

There is of course an intrinsic problem with constructing these soliton solutions numerically. Since we solve the equations of motion
only for a limited range of $\rho$ it is always conceivable that the solution develops pathologies for $\rho \to \pm \infty$.
Below we remedy this deficiency in the most crude way: we plot a region as large as the numerics allows.
This gives already very believable results, and allows to determine the asymptotic behaviour of the solutions.
The next natural steps are to construct these solutions in an asymptotic expansion, that --- as we shall see --- must be non-analytic in $\rho$,
and to perform a stability analysis. We leave these tasks for future work.

With this disclaimer, let us now present a typical example. We choose the data
$\mu = 4$, $\ell = 1$, $\vecX(0) = (-1, 7/3, -2/3)$, $\dot{\vecX}(0) = (-1/5,-1/10,-8/3)$ and $J^+ = -11/4$.
In Fig.~\ref{fig:Geon1} the region near $\rho = 0$ is plotted. $\vecX^2$ has a parabolic shape and never reaches zero. The two 
``constraints'' $\dot{\vecX}\ddot{\vecX}$ and $\ddot{\vecX}^2$ vary substantially and take values of order one, but approach zero as $\rho$ increases. Furthermore, $K$ is clearly non-constant. Most of the interesting features in these plots are localised in a region around $\rho\approx-0.5$.

Zooming out to $|\rho| \approx 10^4$ produces Fig.~\ref{fig:Geon2}. We see that asymptotically
$\dot{\vecX}\ddot{\vecX} \sim 0$ and $\ddot{\vecX}^2 \sim 0$. We therefore expect that the solution asymptotes to a known solution in one of the non-generic sectors.
Inspection of the components of $\ddot{\vecX}$ reveals that they approach constants, indicating that the solution is asymptotically warped AdS. 
Also $K$ seems to approach the exact value $\bar{K}$ [see \eqref{eq:K}] for warped AdS.

While still small, already at these values of $\rho$ the constraint $G$ and equations of motion ${\bf EOM}$ start to deviate from zero due to numerical errors. It turns out that there is a simple fix for this problem. The numerical errors depend heavily on the particular Lorentz frame we use for our computations. Asymptotically, the preferred frame is that in which the acceleration $\ddot{\vecX}$ has only one non-zero component. To improve the numerical accuracy we therefore extract the approximate asymptotic value of $\ddot{\vecX}$ and pass to a frame where $\ddot{Y} = 0$ and $\ddot{X}^- \approx 0$. Of course, we cannot put both exactly zero since $\ddot{\vecX}^2$ is only approximately null. In this new frame numerics is more stable. The whole process can then be iterated to get closer and closer to the optimal frame and to higher and higher values of $\rho$ without compromising the numerical stability.

\begin{figure}[t]
\begin{center}
{\bf Soliton with no centre --- Damped oscillations around warped AdS}\\
\includegraphics[height=140mm]{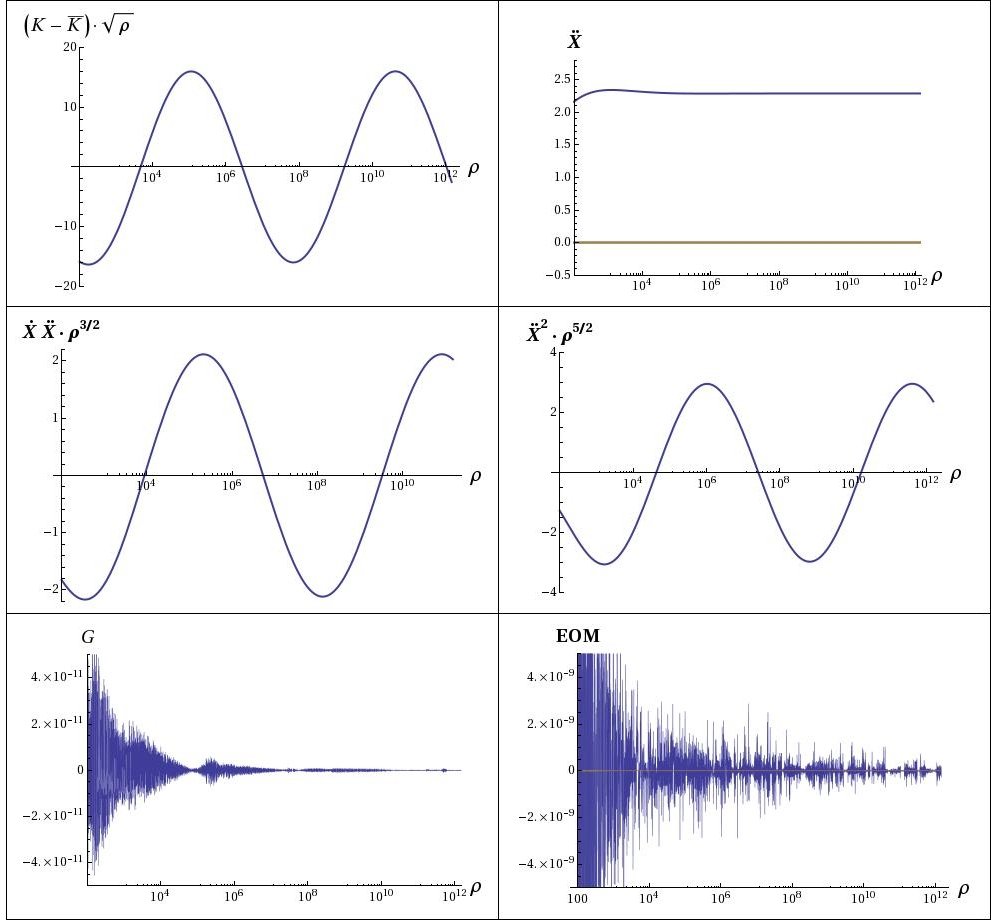}
\caption{This figure shows the same solution as in Fig.~\ref{fig:Geon1} but Lorentz rotated to stabilise the numerics. We see that the numerical errors are small as compared to Fig.~\ref{fig:Geon2}. 
Note that the scale on the $\rho$-axis is logarithmic and that we have multiplied three quantities by half-integer powers of $\rho$.
Asymptotically these quantities exhibit sinusoidal dependence on $\log \rho$. \label{fig:Geon3}}
\end{center}
\end{figure}

In this way it is possible to extend the numerics as far as necessary. We evolve until $|\rho|\sim 10^{12}$. To understand the asymptotics of the solution we
present several quantities plotted in Fig.~\ref{fig:Geon3}. Three of the quantities have been multiplied with suitable functions of $\rho$ to aid the eye in identifying the asymptotic scalings: 
\eq{
(K-\bar{K})\sqrt{\rho}, \qquad \dot{\vecX}\ddot{\vecX}\rho^{3/2}\qquad \mbox{and} \qquad \ddot{\vecX}^2 \rho^{5/2}
}{eq:quantities}
We find that all these quantities to very high accuracy have sinusoidal behaviour in $\log{\rho}$ asymptotically. Therefore we are
led to postulate the following asymptotic behaviour of the soliton solution: 
\eq{
\begin{split}
K - \bar{K} &\sim \rho^{-1/2}\sin\left(\omega \log \rho + \delta_1 \right)\\ 
\dot{\vecX}\ddot{\vecX} & \sim  \rho^{-3/2}\sin\left(\omega \log \rho + \delta_2 \right)\\
 \ddot{\vecX}^2 & \sim \rho^{-5/2}\sin\left(\omega \log \rho + \delta_3 \right)
\end{split}
}{eq:asy1}
for some constants $\omega$ and $\delta_i$.
It is clear, and also explicitly verifiable, that the subleading terms in the components of $\vecX$ also exhibit a similar 
functional behaviour.

\subsection{Generic solution with analytic centre --- Solitons}\label{sec:ex4}

To exemplify the procedure of first Taylor expanding around an analytic centre and then using the
thus obtained values as initial data we present the plots in Figs.~\ref{fig:TnE} and \ref{fig:AnalCenters}.
Both figures illustrate the same solution corresponding to $a=3.7$ and the upper sign in \eqref{eq:as48}, with $\mu = 5$ and $\ell = 1$. The same qualitative behaviour is obtained for the special case obtained when 
setting $a=2/5$ in \eqref{eq:as41}.\footnote{% 
In this special case we use first Algorithm 2 described in \ref{app:alg} for the Taylor expansion. Then we extract initial values at some value of $\rho$ that we feed into Algorithm 3, exactly as described in the text.}

Fig.~\ref{fig:TnE} shows the region close to the centre at $\rho = 0$. Here we include both the Taylor expansion itself and the solution obtained from using the values at $\rho = -0.1$ in the Taylor expansion as initial values. It is clear that the solution is in the generic sector
since neither $\dot{\vecX}\ddot{\vecX}$ nor $\ddot{\vecX}^2$ vanishes.

In Fig.~\ref{fig:AnalCenters} the range of $\rho$ is extended to negative values until the numerics starts to be questionable.
Again, the two quantities $\dot{\vecX}\ddot{\vecX}$ and $\ddot{\vecX}^2$ rapidly approach zero as $\rho \to -\infty$, and $\ddot{\vecX}$ approaches constants.

\begin{figure}[t]
\begin{center}
{\bf Soliton with analytic centre --- Damped oscillations around warped AdS}\\
\includegraphics[height=140mm]{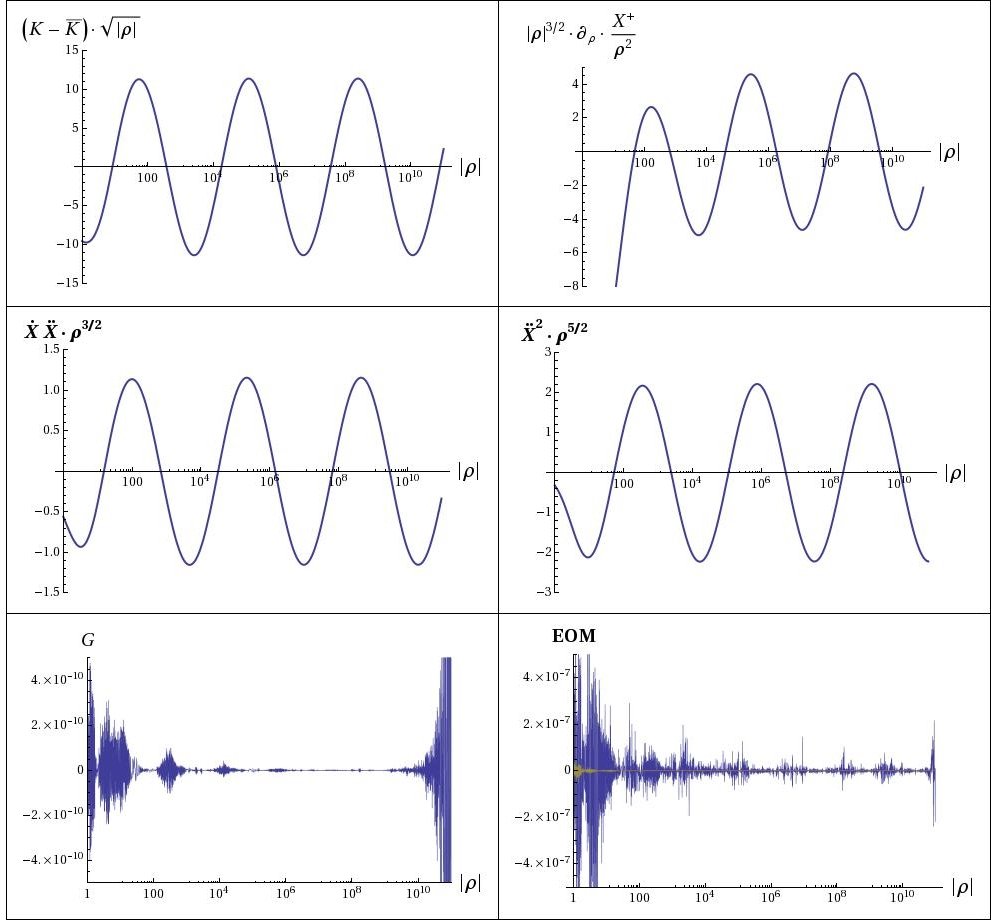}
\caption{The same solution as in Fig.~\ref{fig:AnalCenters} is plotted in a more favourable Lorentz frame. We observe the same oscillating behaviour as in the case without a centre. For this case we also include a plot of $|\rho|^{3/2}\partial_{\rho} (X^+/\rho^2)$ from which the sub-leading behaviour of $X^+$ can be deduced. Note that the errors are small and that we plot against $|\rho| = -\rho$, so the asymptotic region is to the right in the plots. \label{fig:PushedAC}}
\end{center}
\end{figure}

Also for this solution it is possible to deduce the asymptotic behaviour by going to a different Lorentz frame. The result is very similar to the example in subsection \ref{sec:ex3} with exactly the same scalings \eqref{eq:asy1} (but of course with different $\omega$ and $\delta_i$). Note that this implies that the half-integer exponents of $\rho$ in the asymptotic expansion \eqref{eq:asy1} are not functions of $\mu\ell$. This can also be verified by choosing non-integer values for $\mu\ell$. The solution with analytic centre thus is a soliton that asymptotes to warped AdS. We give the corresponding plots in Fig.~\ref{fig:PushedAC}. 

In this example we choose to include a plot (top-right) that reveals the subleading behaviour of $X^+$:
\eq{
X^+ = A \, \rho^2 + B \,\rho^{3/2}\, \sin \left(\tilde{\omega}\log \rho + \tilde{\delta}\right) + \ldots
}{eq:asy2}
Note in particular that the next-to-leading contribution grows faster than $\rho$ which makes the solution incompatible with Comp\`{e}re--Detournay boundary conditions \cite{Compere:2009zj}.

\subsection{Discussion and open questions}\label{sec:disc}

The soliton solutions discovered in section \ref{sec:ex3}, \ref{sec:ex4} asymptote to warped AdS in the sense explained above. At first glance this appears to contradict the no-go result of \cite{Anninos:2010pm}, whose analysis in Appendix E can be generalised straightforwardly to non-null-warped AdS. However, this no-go result relies on the assumption that the solution can be represented as an asymptotic expansion in negative integer powers of $\rho$. Our soliton solutions are capable to circumvent the no-go result since the subleading terms in our asymptotic expressions are not analytic in the coordinate $\rho$ \eqref{eq:asy1}, \eqref{eq:asy2}.

The existence of these soliton solutions raises several further questions that are beyond the scope of the present paper. Let us end this section by listing a few of them.

\bigskip

\begin{itemize}%{$\bullet$}
\item Are there consistent boundary conditions that can 
accommodate the asymptotics \eqref{eq:asy1}, \eqref{eq:asy2}? If so what is the corresponding asymptotic symmetry group?
\item Which boundary terms are needed to yield a vanishing first variation of the action?
\item Are the solutions stable with respect to perturbations that do not respect axi-symmetry and/or stationarity? 
\item Are there any exact solutions of this kind? As the asymptotic functional form is established \eqref{eq:asy1}, \eqref{eq:asy2} it is conceivable that exact solutions can be constructed.
\item Can these solutions be interpreted as finite energy excitations around warped AdS? If this question can be answered affirmative then the soliton solutions found above could have natural interpretations as geons or warped black holes with topologically massive graviton hair.
\item What is the precise range of initial data \eqref{eq:as53}-\eqref{eq:as78} for which we obtain smooth soliton solutions as opposed to naked singularities?  
\item Are there soliton solutions that asymptote to AdS or Schr\"{o}dinger?
\item Are there kink solutions that interpolate between two different non-generic backgrounds, say between AdS and warped AdS? The existence of such solutions could be used to address the stability issues.\footnote{Kink solutions were found in a related 3-dimensional theory, new massive gravity, by Oliva, Tempo and Troncoso \cite{Oliva:2009ip}. Evidence for kink solutions in TMG was reported by Lashkari and Maloney studying the Ricci--Cotton flow in TMM \cite{Lashkari:2010iy}.}
\end{itemize}

\bigskip

\noindent This concludes our presentation of selected numerical solutions of TMM. In summary we have found new soliton solutions in TMG, unravelled their asymptotic warped AdS behaviour and explained why the existence of these solutions does not contradict previous no-go results. We have also raised several questions to be addressed in future work. In the next section we generalise our discussion to non-negative values of the cosmological constant.

\section{Generalisations}\label{sec:gen}

The discussion so far can easily be generalised to vanishing ($\ell\to\infty$) or positive ($\ell^2<0$) cosmological constant. We start with the latter case.

We classify again the solutions into Einstein ($\ddot\vecX=0$), ``Schr\"{o}dinger'' ($\ddot\vecX\neq 0$, $\ddot\vecX=\alpha(\rho)\vecX+\beta(\rho)\dot\vecX$), warped ($\ddot\vecX^2=\dot\vecX\ddot\vecX=0$, $\epsilon_{ijk}\,X^i\dot X^j\ddot X^k\neq 0$) and generic ($\ddot\vecX^2\neq 0$ and/or $\dot\vecX\ddot\vecX\neq0$).

\subsection{Positive cosmological constant}

The essential difference to the AdS case is the sign change of the constant term in the Hamilton constraint \eqref{eq:as5}.

\paragraph{Einstein} The Einstein sector persists, but there are no longer black hole solutions. All solutions are locally dS.

\paragraph{Schr\"{o}dinger} The Schr\"{o}dinger sector does not persist. In other words, there are no solutions where the acceleration $\ddot\vecX$ depends linearly on velocity $\dot\vecX$ and coordinate vector $\vecX$. There is a simple geometric reason for this. Namely, the Hamilton constraint \eqref{eq:as5} requires $\dot\vecX$ to be time-like. On the other hand, the constraints \eqref{eq:as14} must hold and they imply that $\dot\vecX$ is null or space-like. Thus, there are no Schr\"{o}dinger solutions for positive cosmological constant.

\paragraph{Warped} The warped sector persists. In fact, one can take the warped AdS solution \eqref{eq:as40} and just insert negative values of $\ell^2$ to obtain warped dS solutions. 

\paragraph{Generic} One can use here algorithms analog to the ones described in section \ref{sec:5} and \ref{app:alg}.

\subsection{Vanishing cosmological constant}

The essential difference to the AdS case is that the constant term vanishes in the Hamilton constraint \eqref{eq:as5}.

\paragraph{Einstein} Also here the Einstein sector persists, but no black hole solutions arise. All solutions are locally Minkowski.

\paragraph{Schr\"{o}dinger} The ``Schr\"{o}dinger sector'' is modified appreciably as the Hamilton constraint \eqref{eq:as5} implies that $\dot\vecX$ is null. The constraints \eqref{eq:as14} again hold automatically. Up to Lorentz transformations, rescalings and shifts the unique solution is given by
\eq{
\vecX=\big(s e^{-\mu\rho}+a\rho+b,\,0,\,\pm 1\big)
}{eq:as73}
which depends on two integration constants $a,b$ and a sign $s=\pm$. Analog to the AdS-Schr\"{o}dinger case \eqref{eq:as25} we can still exploit $GL(2,\mathbb{R})$ transformations to fix $a,b$ conveniently. The solution \eqref{eq:as73} is well-known \cite{Clement:1992gy}, and it does not exhibit asymptotic Schr\"{o}dinger behaviour. The angular momentum has only one non-vanishing component $J^+=a$.

\paragraph{Warped} The warped sector leads to Vuorio-type of solutions \cite{Vuorio:1985ta}. %,Percacci:1986ja}. 
The solution is obtained from \eqref{eq:as40} in the limit $\ell\to\infty$.

\paragraph{Generic} One can use here algorithms analog to the ones described in section \ref{sec:5} and \ref{app:alg}.
In fact, a numerical study of soliton solutions for this case was performed in \cite{Clement:1992gy} with results similar to those obtained in this work for the cosmological case: Some initial data were found to lead to singularities, while other data lead to smooth solutions.

\subsection{Generalisations beyond topologically massive gravity}

Clement's Ansatz for the line-element \eqref{eq:clement} can be used also for other 3-dimensional theories of gravity, like new massive gravity (NMG) \cite{Bergshoeff:2009hq},\footnote{Indeed, Clement has used this Ansatz recently \cite{Clement:2009gq} %,Clement:2009ka} 
and found warped black hole solutions in NMG.} generalised massive gravity \cite{Bergshoeff:2009aq}, massive supergravity \cite{Andringa:2009yc}, higher order massive gravity \cite{Sinha:2010ai}, extended new massive gravity \cite{Paulos:2010ke} or Born--Infeld massive gravity \cite{Gullu:2010pc}. This could allow to construct stationary axi-symmetric solutions for these novel theories along the lines of our paper: first reduce the gravitational action to a particle mechanics action (see section \ref{sec:2}), next perform a Hamiltonian analysis (see section \ref{sec:3} and \ref{app:A}), then discuss various simple sectors of the theory --- analog to the Einstein, Schr\"{o}dinger\footnote{For NMG, this sector contains both Schr\"{o}dinger and Lifshitz solutions. \cite{AyonBeato:2009yq, AyonBeato:2009nh}.} and warped sectors discussed in the present work (see section \ref{sec:4}) --- and finally provide algorithms to construct generic solutions numerically (see sections \ref{sec:5}, \ref{sec:examples} and \ref{app:alg}). 
It would be interesting to construct soliton solutions for these 3-dimensional theories, to study their stability properties and to address the questions raised at the end of section \ref{sec:disc}.

\section*{Acknowledgements}

We are grateful to Andi Ipp and Roman Jackiw for discussions. 
We thank Eloy Ayon-Beato and Gerard Clement for helpful comments.

SE, DG and NJ are supported by the START project Y435-N16 of the Austrian Science Foundation (FWF) and by the FWF project P21927-N16. NJ acknowledges financial support from the Erwin-Schr\"odinger Institute (ESI) during the workshop ``Gravity in three dimensions'' and financial support from CECS during the Summer workshop on theoretical physics.

\appendix

\section{Poisson brackets and Hamilton equations of motion}\label{app:A}

The non-vanishing Poisson brackets between all twelve constraints \eqref{eq:constraints} read as follows.
\newcommand{\NEG}{\!\!}
\begin{subequations}
\label{eq:app1}
\begin{align}
\NEG \{p^z_i,G\} &= - \Pi_i \approx 0 &
 \{p^z_i,\Phi_j\} &= \frac{1}{\mu}\,\epsilon_{ijk}\,x^k &
 \{p^z_i,x^j\Psi_j\} &= -x_i - \frac{3}{2\mu}\,\epsilon_{ijk}\,x^j y^k \\
\NEG \{G,\Pi_i\} &= \Phi_i \approx 0 &
 \{G,\Phi_j\} &= \Psi_j &
 \{G,x^j\Psi_j\} &= -y^i\Psi_i \\
\NEG \{\Pi_i,\Pi_j\} &= -\frac{1}{\mu}\,\epsilon_{ijk}\,x^k &
 \{\Pi_i,\Phi_j\} &= \eta_{ij} +\frac{1}{2\mu}\, \epsilon_{ijk}\,y^k &
 \{\Pi_i,x^j\Psi_j\} &= \frac{3}{2\mu}\,\epsilon_{ijk}\,x^j z^k \\
\NEG & & \{\Phi_i,\Phi_j\} &= -\frac{2}{\mu}\,\epsilon_{ijk}\,z^k & 
 \{\Phi_i,x^j\Psi_j\} &= -\Psi_i
\end{align}
Here are some additional Poisson brackets of interest.
\begin{align}
 \{p^z_i,y^j\Psi_j\} &= -y_i &\quad
 \{p^z_i,z^j\Psi_j\} &= -2z_i \\
 \{G,y^i\Psi_i\} &= -z^i\Psi_i &\quad
 \{\Pi_i,y^j\Psi_j\} &= -z_i \label{eq:whatever}
\end{align}
All other Poisson brackets between the constraints \eqref{eq:constraints} and the quantities $y^i\Psi_i$, $z^i\Psi_i$ vanish.

The extended Hamiltonian \eqref{eq:as13} with arbitrary $ \la^i$ commutes weakly with all constraints, except with the following ones:
%\begin{align}
%& \{H^{\rm ext},\Phi_i\} \approx e\,\big(\Psi_i-\frac{1}{\mu}\,\epsilon_{ijk}\,\la^jx^k\big) \\
%& \{H^{\rm ext},x^i\Psi_i\} \approx -e\,\big(y^i\Psi_i+\la^i x_i + \frac{3}{2\mu}\,\epsilon_{ijk}\, \la^i x^j y^k\big)
%\end{align}
\eq{
 \{H^{\rm ext},\Phi_i\} \approx e\,\big(\Psi_i-\frac{1}{\mu}\,\epsilon_{ijk}\,\la^jx^k\big) \quad\;\,
 \{H^{\rm ext},x^i\Psi_i\} \approx -e\,\big(y^i\Psi_i+\la^i x_i + \frac{3}{2\mu}\,\epsilon_{ijk}\, \la^i x^j y^k\big)
}{eq:nolabel}
\end{subequations}
Setting to zero the right hand sides in these equations establishes results for the Lagrange multipliers $\la^i$. With the Ansatz
\begin{subequations}
\label{eq:app2}
\eq{
\la^i = \la_x x^i + \la_y y^i + \la_z z^i
}{eq:app1a}
we obtain
%\begin{align}
%& \la_x \, x^2 \approx -y^i\Psi_i - \la_y\,x^iy_i \\
%& \la_y \,\frac1\mu\,\epsilon_{ijk}\,x^iy^jz^k = - z^i\Psi_i \\
%& \la_z \,\frac1\mu\,\epsilon_{ijk}\,x^iy^jz^k =  y^i\Psi_i
%\end{align}
\eq{
 \la_x \, x^2 \approx -y^i\Psi_i - \la_y\,x^iy_i \qquad  \la_y \,\frac1\mu\,\epsilon_{ijk}\,x^iy^jz^k = - z^i\Psi_i \qquad \la_z \,\frac1\mu\,\epsilon_{ijk}\,x^iy^jz^k =  y^i\Psi_i
}{eq:lalapetz}
\end{subequations}
Note that by assumption of regularity $x^2\neq 0$, so $\la_x$ is well-defined. If $x,y,z$ are linearly independent then also $\la_y$ and $\la_z$ are well-defined. If in addition the conditions \eqref{eq:as14} are fulfilled then $\la^i=0$ and the extended Hamiltonian $H^{\rm ext}$ reduces to the canonical one $H=e\,G$. The Poisson brackets between the canonical Hamiltonian and the conditions \eqref{eq:as14} either vanish or yield something proportional to these conditions, see \eqref{eq:whatever}. Therefore, if the  conditions \eqref{eq:as14} both hold initially they must hold for all times. If $x,y,z$ are linearly dependent then the constraints \eqref{eq:as14} are fulfilled automatically, as discussed in the main text in section \ref{sec:L}.

The Hamilton equations of motion generated by the extended Hamiltonian \eqref{eq:as13} are
\begin{subequations}
\label{eq:Heom}
\begin{align}
\dot e &= 0 & \dot p^e &\approx 0 \\
\dot x^i &= e\, y^i & \dot p^x_i &\approx -\frac{e}{2\mu}\,\epsilon_{ijk}\,y^j z^k \\ 
\dot y^i &= e\, z^i & \dot p^y_i &\approx -\frac{e}{2\mu}\,\epsilon_{ijk}\,x^j z^k\\
\dot z^i &= e\, \la^i & \dot p^z_i &\approx 0 
\end{align}
\end{subequations} 
We see that the choice \eqref{eq:as13} for the extended Hamiltonian automatically leads to the desired gauge condition $e=\rm const.$ With no loss of generality we set $e=1$ throughout in this work.

\section{Proof that all higher order polynomial solutions are Schr\"{o}dinger}\label{app:proof}

To show the absence of generic solutions that are polynomial in the radial coordinate $\rho$ we insert the Taylor series $\vecX=\sum_{n=0}^N\vecX_{(n)}\,\rho^n$ into the equations of motion \eqref{eq:as6} and assume $N\geq 3$, since for $N\leq 2$ we recover non-generic solutions.
\begin{multline}
\sum_{n=2}^N n(n-1)X_{(n)\,i}\,\rho^{n-2} = -\frac{1}{2\mu}\,\epsilon_{ijk}\,\Big(3\sum_{n=1}^N\sum_{m=2}^N nm(m-1) X_{(n)}^j X_{(m)}^k\, \rho^{n+m-3}\\
+2\sum_{n=0}^N\sum_{m=3}^N m(m-1)(m-2) X_{(n)}^j X_{(m)}^k\, \rho^{n+m-3}\Big)
\label{eq:proof1}
\end{multline}
We then solve \eqref{eq:proof1} order by order in $\rho$, starting with the highest power $\rho^{2N-3}$. We obtain the following chain of conditions:
\begin{align}
\rho^{2N-3}: \qquad & \textrm{holds\;trivially} \\
\rho^{2N-4}: \qquad & \epsilon_{ijk}\,X_{(N-1)}^j X_{(N)}^k = 0 \\
\rho^{2N-5}: \qquad & \epsilon_{ijk}\,X_{(N-2)}^j X_{(N)}^k = 0 \\
\rho^{2N-6}: \qquad & \epsilon_{ijk}\,\big(X_{(N-3)}^j X_{(N)}^k+\frac{2N-4}{3N}\, X_{(N-2)}^j X_{(N-1)}^k \big) = 0 \\
\dots \nonumber
\end{align}
and similar relations down to the power of $\rho^{N-1}$. These relations involve only the right hand side of the equations of motion \eqref{eq:proof1} and taken together they imply a chain of proportionalities
\eq{
\vecX_{(n)}\propto\vecX_{(N)} \qquad \forall n\geq 2
}{eq:proof2}
Apart from the proportionality constants in \eqref{eq:proof2} only three coefficient vectors are free at this stage: $\vecX_{(0)}$, $\vecX_{(1)}$ and $\vecX_{(N)}$. We then continue to solve \eqref{eq:proof1}  order by order in $\rho$, starting with the power $\rho^{N-2}$ down to the power $\rho^0$.  We obtain the following chain of conditions:
\begin{align}
\rho^{N-2}: \qquad & X_{(N)\,i} = -\frac{2N-1}{2\mu}\,\epsilon_{ijk}\,X_{(1)}^jX_{(N)}^k \\
\rho^{N-3}: \qquad & X_{(N-1)\,i} = -\frac{1}{2\mu}\,\epsilon_{ijk}\,\Big((2N-3)X_{(1)}^jX_{(N-1)}^k +2NX_{(0)}^jX_{(N)}^k\Big)\\
\dots \nonumber \\
\rho^0: \qquad & X_{(2)\,i} = -\frac{1}{2\mu}\,\epsilon_{ijk}\,\Big(3X_{(1)}^j X_{(2)}^k+6X_{(0)}^jX_{(3)}^k\Big)
\end{align}
These relations taken together imply a chain of inner products
\eq{
\vecX_{(1)}\vecX_{(n)} = \vecX_{(n)}\vecX_{(n)} = 0\qquad \forall n\geq 2
}{eq:proof3}
and the vanishing of all triple products
\eq{
\epsilon_{ijk}\, X_{(0)}^i X_{(1)}^j X_{(n)}^k = 0\qquad \forall n
}{eq:proof4}
Therefore, the acceleration $\ddot\vecX$ must depend linearly on the vector $\vecX$ and its velocity $\dot\vecX$. This is the defining feature of a Schr\"{o}dinger solution, so all polynomial solutions of at least third order in $\rho$ are Schr\"{o}dinger solutions. This is what we wanted to prove.

\section{Solutions for special values of the coupling constant}\label{app:D}

In this appendix we collect some interesting features of solutions for special values of the dimensionless coupling constant $|\mu\ell|$. We always keep $\ell$ finite and positive, even when taking certain singular limits. We start with the (singular) limit $|\mu\ell|\to 0$. Then we address the logarithmic point ($|\mu\ell|=1$), the null warped point ($|\mu\ell|=3$) and the special point $|\mu\ell|=5$. Finally we discuss the (singular) limit $|\mu\ell|\to\infty$.

\subsection{Chern--Simons point $|\mu\ell|\to 0$}\label{sec:ex5}

In this (singular) limit only the gravitational Chern--Simons term remains in the action \eqref{eq:intro1}. The limiting theory does not exhibit any propagating physical degree of freedom. The equations of motion require the Cotton tensor to vanish. Thus any solution of the limiting theory that is also a solution of TMG must belong to the Einstein sector. We shall exhibit this feature explicitly below.

All solutions of the limiting theory are known in closed form. They can be obtained as follows. Take the gravitational Chern--Simons term and perform a Kaluza--Klein reduction to two dimensions \cite{Guralnik:2003we} along the angular Killing direction. 
\eq{
g_{\mu\nu}=\varphi\,\left(\begin{array}{cc}
g_{\al\be}-a_\al a_\be & -a_\al \\
-a_\be & -1
\end{array}\right)_{\mu\nu}
}{eq:as84}
Here $g_{\al\be}$ is a 2-dimensional metric and $a_\al$ a 2-dimensional gauge field. The scalar field that appears in the Kaluza--Klein Ansatz \eqref{eq:as84} does not play any role in the reduced theory and is set to unity for the time being, $\varphi=1$. This reduction yields a specific Maxwell-dilaton gravity model, the solutions of which can be obtained locally and globally in closed form \cite{Grumiller:2003ad}. There are some isolated solutions, so-called constant dilaton vacua, and a set of generic solutions. 

In an Eddington--Finkelstein patch the generic solution reads \cite{Grumiller:2003ad}
\eq{
\extd s^2_{\textrm{\tiny 2D}}=g_{\al\be}\extd x^\al\extd x^\be = 2\extd u \extd\rho + \big(2{\cal C}-\frac c2\,\rho^2+\frac14\,\rho^4\big)\extd u^2\qquad a_\al = \frac12\,\rho^2\,\de_\al^u
}{eq:as85}
The quantities $c,{\cal C}$ are constants of motion and have a 2-dimensional interpretation as conserved $U(1)$-charge and conserved mass, respectively. For BPS saturation \cite{Bergamin:2004me}, $c^2=8{\cal C}$, the solution \eqref{eq:as85} is the kink solution discovered in \cite{Guralnik:2003we}. Plugging the 2-dimensional solution \eqref{eq:as85} into the Kaluza--Klein Ansatz \eqref{eq:as84} yields the 3-dimensional line-element
\eq{
\extd s^2=2\extd u\extd\rho+\big(2{\cal C}-\frac c2\,\rho^2\big)\,\extd u^2-\rho^2\,\extd u\extd\phi-\extd\phi^2
}{eq:as86} 
The geometry defined by the line-element \eqref{eq:as86} has vanishing Cotton-tensor and thus is conformally flat, but it has non-vanishing Ricci-tensor. The corresponding Ricci-scalar is not constant, $R=c-\frac52\,\rho^2$. Thus the solution \eqref{eq:as86} cannot arise directly as the smooth limit of some TMG solutions for finite values of $|\mu\ell|$. However, for non-vanishing mass, ${\cal C}\neq 0$, we can multiply the solution \eqref{eq:as86} with the conformal factor $\varphi=2{\cal C}\ell^2/\rho^2$ so that the transformed Ricci-scalar is constant and negative, $\widetilde R = -6/\ell^2$. This solution then arises as the smooth limit of Einstein sector TMG solutions for finite values of $|\mu\ell|$.  

The 2-dimensional equations of motion exhibit a $\mathbb{Z}_2$-symmetry $a_\al\to-a_\al$ \cite{Guralnik:2003we}. One of the isolated solutions preserves this $\mathbb{Z}_2$-symmetry
\eq{
\extd s^2=2\extd u\extd\rho+\big(A+B\rho-\frac c2\,\rho^2 \big)\,\extd u^2-\extd\phi^2
}{eq:as87}
while the others break it
\eq{
\extd s^2=2\extd u\extd\rho+\big(A+B\rho\big)\,\extd u^2\pm 2\sqrt{c}\rho\,\extd u\extd\phi-\extd\phi^2
}{eq:as88}
In all cases $A,B,c$ are constants of motion, the Ricci-scalar is constant and the Cotton-tensor vanishes. For $c=4/\ell^2$ the $\mathbb{Z}_2$-symmetry breaking solutions \eqref{eq:as88} directly arise as the smooth limit of Einstein sector TMG solutions for finite values of $|\mu\ell|$. For other values of $c\neq 0$ we have to multiply the line-element \eqref{eq:as88} by a constant conformal factor $\varphi=c\ell^2/4$ in order to obtain the appropriate transformed Ricci-scalar $\widetilde R = -6/\ell^2$. Similarly, we have to multiply the  $\mathbb{Z}_2$-symmetry preserving line-element \eqref{eq:as87} by a conformal factor $\varphi=\ell^2(A^2+2Bc)/(2c(\rho-A/c)^2)$ in order to obtain the appropriate transformed Ricci-scalar $\widetilde R = -6/\ell^2$. This is possible as long as $c\neq 0$ and $A^2+2Bc\neq 0$.

In conclusion, the solutions of TMG at $\mu\ell = 0$ form classes of conformally related metrics. In nearly all such classes
there is a representative that can arise as the smooth limit of Einstein sector TMG solutions for finite values of $|\mu\ell|$. 

\subsection{Logarithmic point $|\mu\ell|=1$}\label{sec:ex6}

At this point Einstein solutions necessarily have light-like angular momentum \eqref{eq:as37}. Schr\"{o}dinger solutions have a logarithmic term in $\rho$ and can either be asymptotically AdS \eqref{eq:as26} [compare with \eqref{eq:as30}] or marginally violate the condition of asymptotically AdS \eqref{eq:as27}. Neither warped nor generic solutions do exhibit special features at this point.

\subsection{Null warped point $|\mu\ell|=3$}\label{sec:ex7}

Schr\"{o}dinger solutions that are not asymptotically AdS have a scaling exponent $z=2$ \eqref{eq:as32}. The coordinate vector $\vecX$ is quadratic in the coordinate $\rho$ \eqref{eq:as25} for these solutions. The angular momentum is light-like\eqref{eq:as28} [see also \eqref{eq:as50}]. The null warped solution coincides with the Schr\"{o}dinger solution. This is the most distinguishing feature of the null warped point. Neither Einstein nor generic solutions do exhibit special features at this point.

\subsection{Special point $|\mu\ell|=5$}\label{sec:ex8}

Generic solutions with analytic centre can have peculiar properties. Namely, the conditions \eqref{eq:as83} simultaneously can hold at $\rho=0$. We stress that this is the only value for the dimensionless coupling constant $|\mu\ell|$ where generic solutions of this type can exist. For any other value of $|\mu\ell|$ the constraints \eqref{eq:as83} imply the conditions \eqref{eq:as14} and thus do not lead to solutions of the generic sector. The algorithm to construct these solutions differs from the algorithms that are used at any other value of the coupling constant, see Algorithm 2 in \ref{app:alg}. Neither Einstein, nor Schr\"{o}dinger nor warped solutions do exhibit special features at this point.

\subsection{Einstein point $|\mu\ell|\to\infty$}\label{sec:ex9}

In this (singular) limit only the Einstein--Hilbert term and the cosmological constant remain in the action \eqref{eq:intro1}. The limiting theory does not exhibit any propagating physical degree of freedom. Clearly, all solutions in this limit are in the Einstein sector and therefore locally and asymptotically AdS.

\section{Algorithms}\label{app:alg}

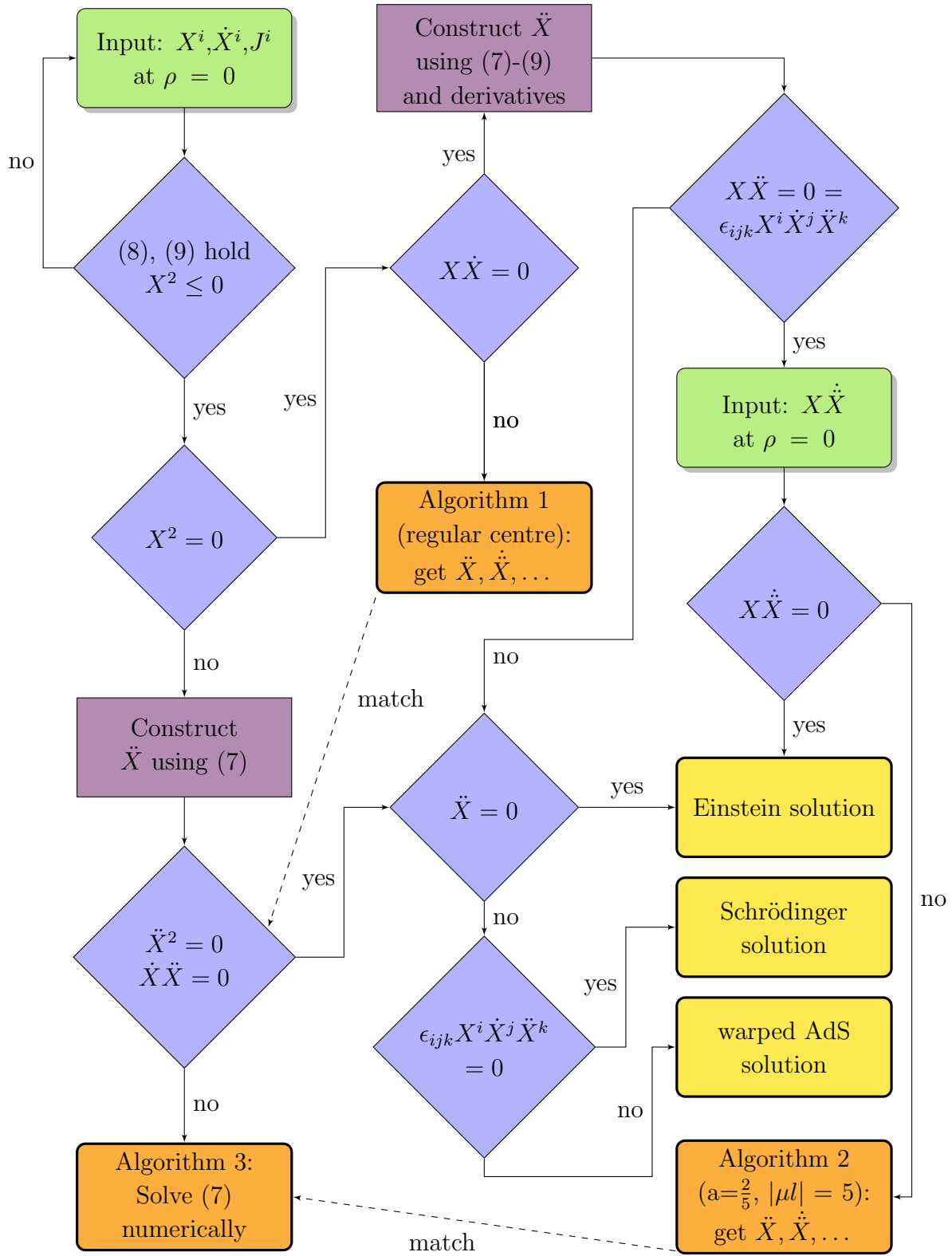
\begin{figure}[ht]
\begin{center}
% Define block styles
\tikzstyle{decision} = [diamond, draw, fill=blue!30, text width=6em, text badly centered, node distance=3cm, inner sep=2pt]
\tikzstyle{block} = [rectangle, draw, fill=taplum!90, text width=8em, text centered, minimum height=4em]
\tikzstyle{line} = [draw, -latex']
\tikzstyle{endblock} = [rectangle, draw, fill=taorange, text width=8em, text centered, rounded corners, minimum height=4em]
\tikzstyle{inblock} = [rectangle, draw, fill=tachameleon!60, text width=8em, text centered, rounded corners, minimum height=4em, drop shadow]
\tikzstyle{solblock} = [rectangle, draw, fill=tabutter, text width=8em, text centered, rounded corners, minimum height=4em]
  
\begin{tikzpicture}[node distance = 3cm,auto]
    % Place nodes
    \node [inblock] (input) {Input: $X^i$,$\dot{X}^i$,$J^i$ at $\rho=0$};
    \node [decision, below of=input, node distance=3.5cm] (hold) {\eqref{eq:as8}, \eqref{eq:as9} hold $X^2\leq 0$};
    \node [decision, below of=hold, node distance=4.5cm] (X2) {$X^2=0$};
    \node [decision, right of=hold,node distance=5cm] (XXdot) {$X\dot{X}=0$};
    \node [block, below of=X2,node distance=3.5cm] (construct7) {Construct $\ddot{X}$ using \eqref{eq:as7}};
    \node [decision, below of=construct7,node distance=3.5cm] (XDdot2) {$\ddot{X}^2=0$ $\dot{X}\ddot{X}=0$};
    \node [endblock, below of=XDdot2,node distance=4cm,very thick] (generic3) {Algorithm~3: Solve \eqref{eq:as7} numerically};
    \node [endblock, below of=XXdot, node distance=4.5cm,very thick] (generic2) {Algorithm~1 (regular centre): get $\ddot{X},\dot{\ddot{X}},\dots$};
    \node [decision, below of=generic2, node distance=4.5cm] (XDdot){$\ddot{X}=0$};
    \node [decision, below of=XDdot, node distance=4cm] (epsilon1){$\epsilon_{ijk}X^i\dot{X}^j\ddot{X}^k$  $=0$}; 
    \node [solblock,right of=XDdot, node distance=5cm,very thick] (einstein){Einstein solution};
    \node [solblock, right of=epsilon1, node distance=5cm,very thick] (wads){warped AdS solution};
    \node [block, right of=input, node distance=5cm] (construct XDdot){Construct $\ddot{X}$ using \eqref{eq:as7}-\eqref{eq:as9} and derivatives};
    \node [decision, above of=einstein, node distance=10cm] (epsilon2){$X\ddot{X}=0$ $=\epsilon_{ijk}X^i\dot{X}^j\ddot{X}^k$};
    \node [inblock, below of=epsilon2, node distance=3.5cm] (newb){Input: $X\dtdot{X}$ at $\rho=0$};
    \node [decision, below of=newb, node distance=3.1cm] (XXtdot){$X\dot{\ddot{X}}=0$};
    \node [endblock, right of=generic3, node distance=10cm,very thick] (generic1){Algorithm~2 (a=$\frac{2}{5}$, $|\mu l|=5$): get $\ddot{X},\dot{\ddot{X}},\dots$};
    \node [solblock, below of=einstein, node distance=2cm,very thick] (lif){Schr\"{o}dinger solution};
% Draw edges
    \path [line] (input) -- (hold);
    \draw [line] (hold.west) -- ++ (-0.5,0) |-  node [near start]{no} (input.west);
    \path [line] (hold) -- node {yes}(X2);
    \path [line] (X2) -- node {no} (construct7);
    \path [line] (construct7) -- (XDdot2);
    \path [line] (XDdot2) -- node {no}(generic3);
    \path [line] (X2.east) -- ++ (0.8,0) |-  node [near start] {yes}(XXdot);
    \draw [line] (XXdot)-- node {no} (generic2);
    \path [line,dashed] (generic2.west)+(0,-1.0) -- node [near start]{match}(1.4,-14.5); %(XDdot2);
    \path [line,dashed] (generic1.west)+(0,-0.9) -- node {match}(generic3.east);
    \path [line] (XDdot2.east) -- ++ (0.8,0) |- node [near start]{yes} (XDdot);
    \path [line] (XDdot) --node {no} (epsilon1);
    \path [line] (epsilon1.south) -- ++ (0,-0.3) -| ++ (2.8,0) |- node [near start]{no}(wads.west); 
    \path [line] (XXdot) -- node [near start] {yes}  (construct XDdot); 
    \path [line] (construct XDdot.east) -- ++ (0,0) -| (epsilon2.north);
    \path [line] (epsilon2.west) -- ++ (-0.6,0) |- (5,-9.7) -- node [near start]{no}(XDdot.north); 
    \path [line] (epsilon2) -- node {yes}(newb); 
    \path [line] (newb) -- (XXtdot);
    \path [line] (XDdot) -- node {yes} (einstein); 
    \path [line] (XXtdot) -- node {yes} (einstein);
    \path [line] (XXtdot.east) -- ++ (0.5,0) |- node [near start]{no} (generic1); 
    \path [line] (epsilon1.east) -- ++ (0.5,0) |- node [near start] {yes} (lif.west);
    \path [line] (XXdot) -- node {no} (generic2);
\end{tikzpicture}
\caption{Flow chart for the decision which algorithm to use starts in upper left corner}
\label{fig:1}
\end{center}
\end{figure}

In Fig.~\ref{fig:1} we depict a flow chart for the decision which of the algorithms below should be used (if any), given some initial data $\vecX$, $\dot\vecX$ at $\rho=0$ and the conserved angular momentum $\vecJ$. For instance, if the constraints \eqref{eq:as14} both hold then no algorithm is needed since we have a non-generic solution, that is, either Einstein or Schr\"{o}dinger or warped. These solutions are all known analytically. If the constraints \eqref{eq:as14} do not hold then for random initial data one can use the Algorithm 3 below. If the initial data are fine-tuned so that a regular centre emerges and the starting point of the evolution coincides with the centre then one has to use Algorithms 1 or 2, depending on the outcome of the flow chart. The matching procedure indicated by the dashed arrows works as follows: we follow Algorithms 1 or 2 up to the required level of iteration, say, hundred steps. In this way we know the value of the first hundred derivatives of $\vecX$ at $\rho=0$ and thus we can construct the solution at some finite value of $\rho$ for a given accuracy, say, at $\rho=0.01$. We then take the values of $\vecX, \dot\vecX$ at $\rho=0.01$ and the values of $\vecJ$ as new initial data and feed them into Algorithm 3.

In the algorithms below we assume that Lorentz transformations and rescalings have been exploited as explained in the main text in order to reduce the amount of independent initial data. However, we stress that the flow chart in Fig.~\ref{fig:1} works for any set of initial data, and that the algorithms below easily can be adapted to initial data that has not been reduced by these transformations.

\algorithm{Algorithm 1 --- $\vecX^2=0$, $\vecX\dot\vecX\neq 0$ at $\rho=0$}

\paragraph{Step 1: Initial data} One has to choose the upper or lower sign in the Ansatz \eqref{eq:as48} and a value for $a$ that differs from the values listed in \eqref{eq:as51} and is compatible with $\vecX$ being spacelike for $\rho>0$. 
\eq{
\pm\mu<0\qquad\qquad a \neq  \pm 4\, \big(1-\frac{1}{\mu^2\ell^2}\big) \qquad\qquad a\neq\pm\frac{32}{9}
}{eq:aineq}
The initial data \eqref{eq:as48} by construction are compatible with the constraints \eqref{eq:as7}-\eqref{eq:as9}.

\paragraph{Step 2: Solving for $\ddot\vecX$} Since the matrix \eqref{eq:as46} is invertible we obtain a unique solution for $\ddot\vecX$ at $\rho=0$ by solving the linear equations \eqref{eq:as45}.

\paragraph{Step 3: Solving for $\dtdot{\vecX}$} Similarly, we obtain a unique solution for $\dtdot{\vecX}$ at $\rho=0$ by solving the linear equations \eqref{eq:as47}. Note that the matrix appearing on the left hand side of \eqref{eq:as47} necessarily is invertible.

\paragraph{Step $n$: Solving for $\extd^n\vecX/\extd\rho^n$} %$\overset{(n)}\vecX$} 
By taking the $n$-th $\rho$-derivative of \eqref{eq:as43} and evaluating it at $\rho=0$ we obtain three linear equations that allow to determine $\overset{(n)}\vecX$ at $\rho=0$. The crucial input here is that all the matrices appearing in this construction are invertible,
\eq{
A_{ik}^{(n)}:=A_{ik}-\frac{2(n-2)}{\mu}\,\vecX\dot\vecX\eta_{ik}\qquad \det A^{(n)}\neq 0\; \forall\; n\geq 2
}{eq:as75}
with $A_{ik}$ as defined in \eqref{eq:as46}. We obtain the evolution equation
\eq{
A_{ik}^{(n)}\,\overset{(n)}{X^k} = \frac{9}{8\mu}\,\frac{\extd^{n-2}}{\extd\rho^{n-2}}\Big(\big(\dot\vecX^2+\frac{4}{\ell^2}\big)\dot X_i\Big) - \sum_{m=1}^{n-2}\frac{\extd^{m-1}}{\extd\rho^{m-1}}\,\Big(\dot A_{ik}^{(n-m)}\,\overset{(n-m)}{X^k}\Big)\qquad n\geq 3
}{eq:as76}
The right hand side of \eqref{eq:as76} contains $\vecX$ and its first $n-1$ derivatives, all of which can be determined iteratively by using \eqref{eq:as76} repeatedly. The matrix on the left hand side of \eqref{eq:as76} is sensitive to the initial data \eqref{eq:as48}, only.

\paragraph{Final Step: Summing the Taylor series} Given the solutions for $\vecX$, $\dot\vecX$, $\ddot\vecX$, ... $\overset{(n)}\vecX$ at $\rho=0$ for some finite $n$ the solution for $\vecX$ is approximately given by
\eq{
\vecX \approx \vecX\big|_{\rho=0} + \rho \dot\vecX\big|_{\rho=0} + \frac{\rho^2}{2} \ddot\vecX\big|_{\rho=0}  + \dots + \frac{\rho^n}{n!}\,\overset{(n)}\vecX\big|_{\rho=0} 
}{eq:as74} 
provided that $\rho$ is sufficiently small. At some finite value of $\rho$ one can then match to the algorithm for solutions with no analytic centre described below to evolve to arbitrary values of $\rho$.

\begin{center}
* * *
\end{center}

\algorithm{Algorithm 2 --- $\vecX^2=\vecX\dot\vecX=0$ at $\rho=0$}
\paragraph{Step 1: Initial data} This algorithm only is applicable when $|\mu\ell|=5$. One has to choose $s=\pm 1,0$ in the Ansatz \eqref{eq:as41} and $a=2/5$. A convenient choice for the free initial datum is the value of $\dtdot{X}^-$ at $\rho=0$.

\paragraph{Step 2: Consistency of $\ddot\vecX$} The condition \eqref{eq:as45} simplifies to $A_{ik}\ddot X^k=0$, where the matrix $A_{ik}$ now has rank 1 only. This condition then establishes $\ddot X^-=0$.

\paragraph{Step 3: Solving for $\ddot\vecX$} Taking the $\rho$-derivative of \eqref{eq:as43} yields 
\eq{
B_{ik} \ddot X^k = A_{ik}\dtdot{X}^k 
}{eq:as79}
with the invertible matrix
\eq{
B_{ik} = -\frac{6}{\mu\ell^2}\,\eta_{ik}+\frac{19}{4\mu}\,\dot X_i \dot X_k + \frac{5}{2\mu}\,X_i\ddot X_k - \epsilon_{ijk}\,\dot X^j
}{eq:as80}
The quadratic set of equations \eqref{eq:as79} can be solved uniquely for $\ddot X^k$ in terms of the initial datum $\dtdot{X}^-$. It is better to solve first the $Y$-component of \eqref{eq:as80} for $\ddot X^Y$ and then the minus component of \eqref{eq:as80} for $\ddot X^+$. In this way only linear equations have to be solved. The plus component of \eqref{eq:as80} is an identity so that the solution for $\ddot\vecX$ necessarily is compatible with the consistency requirement of Step 2 above.

\paragraph{Step $n+1$: Solving for $\extd^n\vecX/\extd\rho^n$} %$\overset{(n)}\vecX$} 
By taking the $n$-th $\rho$-derivative of \eqref{eq:as43} and evaluating it at $\rho=0$ we obtain three linear equations that allow to determine $\overset{(n)}\vecX$ at $\rho=0$. The crucial input here is that all the matrices appearing in this construction are invertible,
\eq{
B_{ik}^{(n)} = \frac{9}{4\mu}\,\dot X_i\dot X_k+\frac{5}{2\mu}\,X_i\ddot X_k-\sum_{m=2}^n \dot A_{ik}^{(m)}\qquad \det B^{(n)}\neq 0\; \forall\; n\geq 3
}{eq:as81}
where the matrices $A^{(n)}$ are defined in \eqref{eq:as75} below. We obtain the evolution equation
\begin{multline}
-B_{ik}^{(n)}\,\overset{(n)}{X^k} + A_{ik}\,\overset{(n+1)}{X^k} = \frac{9}{8\mu}\,\frac{\extd^{n-1}}{\extd\rho^{n-1}}\Big(\big(\dot\vecX^2+\frac{4}{\ell^2}\big)\dot X_i\Big) -\frac{9}{4\mu}\,\overset{(n)}\vecX \dot\vecX\, \dot X_i -\frac{5}{2\mu}\,\overset{(n)}\vecX\ddot\vecX X_i \\
- \sum_{m=1}^{n-2}\frac{\extd^{m}}{\extd\rho^{m}}\,\Big(\dot A_{ik}^{(n-m)}\,\overset{(n-m)}{X^k}\Big)+\sum_{m=2}^{n-1} \dot A_{ik}^{(m)}\overset{(n)}{X^k}\qquad n\geq 3
\label{eq:as82}
\end{multline}
 The right hand side of the evolution equation \eqref{eq:as82} contains up to $n-1$ derivatives of $\vecX$ (note in particular that all $n$-derivative terms cancel). The left hand side  of the evolution equation \eqref{eq:as82} does contain not only $\overset{(n)}\vecX$, but the minus component of this equation also is sensitive to $\overset{(n+1)}\vecX$. This component can then be determined as follows: we solve \eqref{eq:as82} for $\overset{(n)}\vecX$ in terms of lower derivatives (which are known at this Step) and the minus component of $\overset{(n+1)}\vecX$. Since the minus component of $\vecX$ is determined from Step $n$ already, we obtain in this way one equation for the unknown minus component of $\overset{(n+1)}\vecX$.

\paragraph{Final Step: Summing the Taylor series} This step is exactly the same as in Algorithm 1 above.

\begin{center}
* * *
\end{center}

\algorithm{Algorithm 3 --- $\vecX^2\neq 0$ at $\rho=0$}

\paragraph{Step 1: Initial data} One chooses three constants $a,b, c^+$ and starts with the initial data \eqref{eq:as53} or \eqref{eq:as58} if $b\neq 0$. By construction they are compatible with the constraints \eqref{eq:as8}-\eqref{eq:as9}. If $b=0$ but $a\neq 0$ then one takes instead the initial data \eqref{eq:as77} or \eqref{eq:as78}. If $a=b=0$ then one does not obtain a solution of the generic sector but rather an Einstein solution.

\paragraph{Final step: Time evolution} With these initial data one solves numerically the coupled set of second order ordinary differential equations \eqref{eq:as7}. This can be done straightforwardly with some computer algebra system. As long as $\vecX^2\neq 0$ the configurations obtained in this way will remain regular. If there is a value for $\rho$ where $\vecX^2\to 0$ then numerics will break down eventually and one cannot follow the evolution beyond this point. But such a locus anyhow is a boundary of our coordinate patch. It is worthwhile emphasising that any solution obtained in this way automatically solves the original equations of motion \eqref{eq:as6} {\em and} the Hamilton constraint \eqref{eq:as5}.

\begin{center}
* * *
\end{center}

\section*{References}

%\bibliographystyle{fullsort}
%\bibliography{review} 

\providecommand{\href}[2]{#2}\begingroup\raggedright\endgroup

\clearpage

\begin{figure}[H]
\begin{center}
\includegraphics[height=180mm]{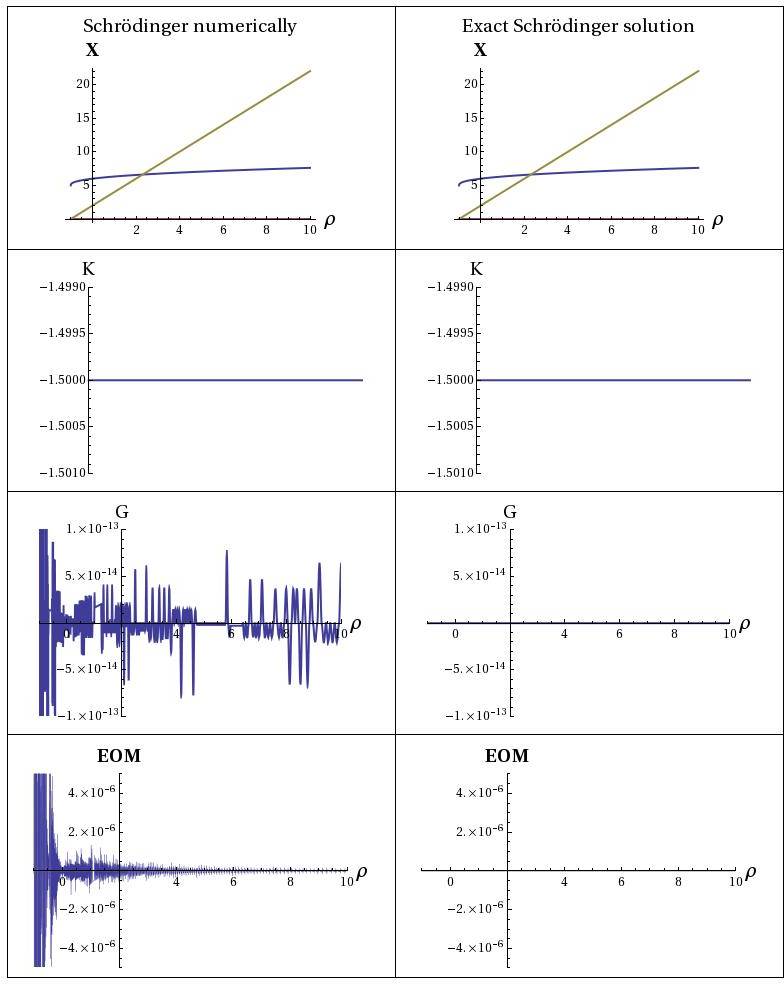}
\caption{Plotted side by side are $\vecX$, $K$, $G$ and {\bf EOM} for a Schr\"{o}dinger solution. The left column of plots was obtained numerically, whereas the right column corresponds to the exact solution. This particular solution corresponds to the values $\mu = 0.2$, $\ell = 1$, $a=0$, $b = 5$ and $s = +1$ in \eqref{eq:as25}. We start evolving from $\rho=1$, which is shifted to $\rho=0$ in the plots.\label{fig:L}} 
\end{center}
\end{figure}

\begin{figure}[H]
\begin{center}
\includegraphics[height=180mm]{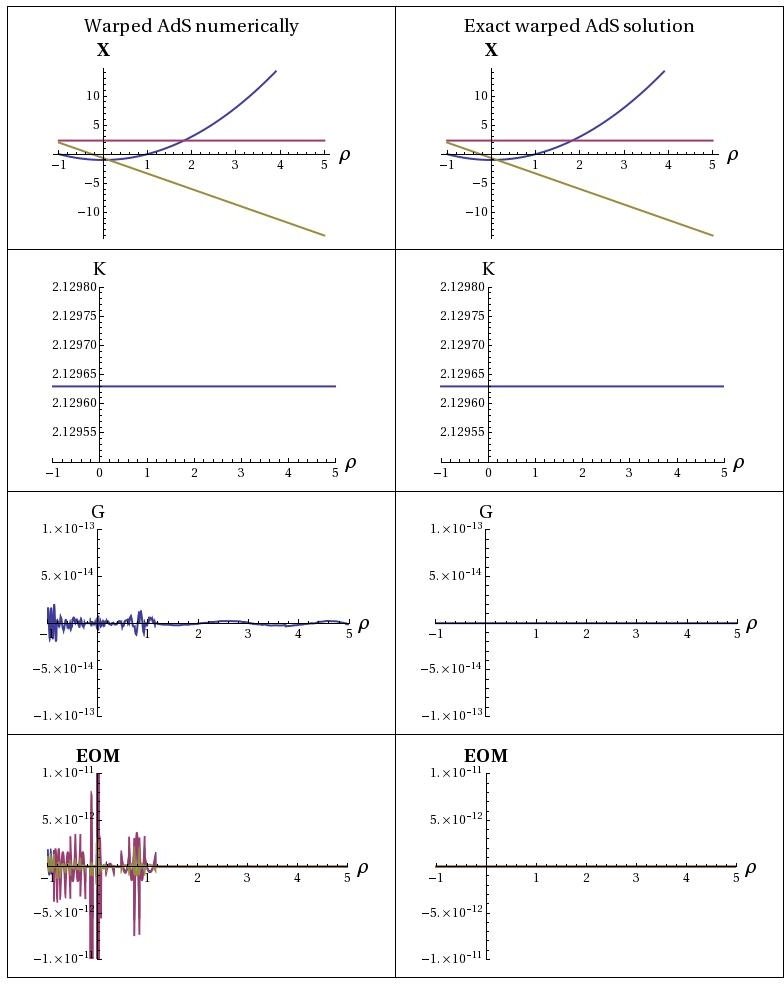}
\caption{Plotted side by side are $\vecX$, $K$, $G$ and {\bf EOM} for a warped AdS solution. The left column of plots was obtained numerically, whereas the right column corresponds to the exact solution. This particular solution corresponds to the values $\mu = 4$, $\ell = 1$ and $a=2$ in \eqref{eq:as40}. We start evolving from $\rho=1$, which is shifted to $\rho=0$ in the plots.\label{fig:W}}
\end{center}
\end{figure}

\begin{figure}[H]
\begin{center}
{\bf Naked singularity solution}\\
\includegraphics[height=180mm]{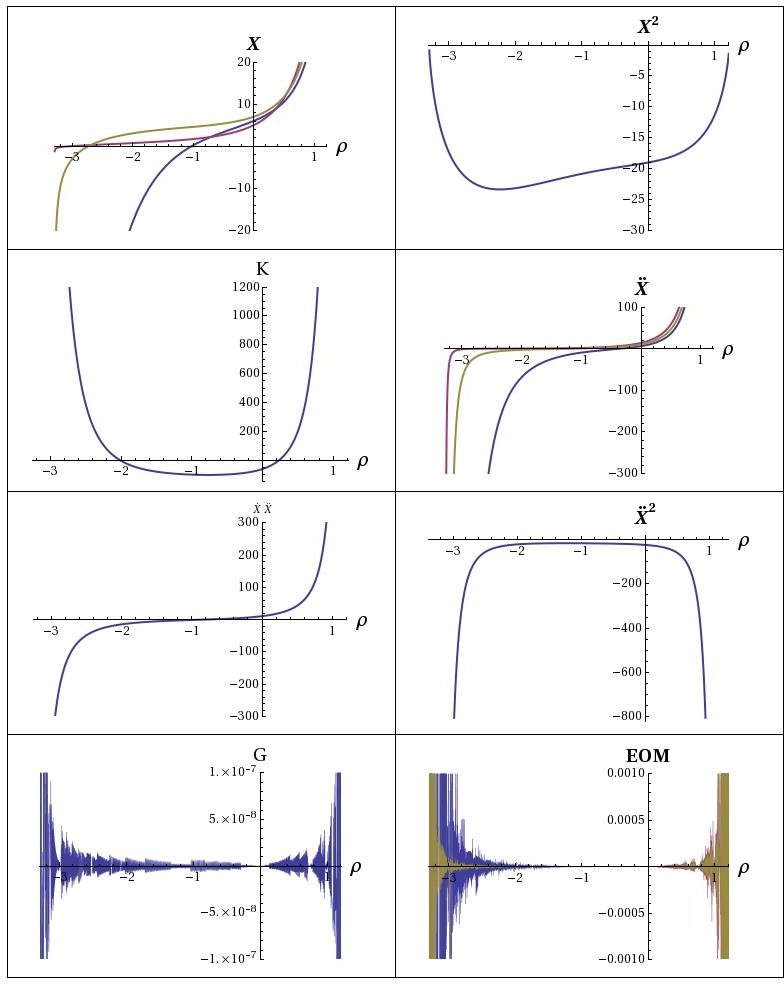}
\caption{The plots show a numerical solution in the generic sector. It has two non-analytic centres at $\rho \approx -3.3$
and $\rho \approx 1.21$ respectively. As can be deduced from the plot of $K$ both centres correspond to curvature singularities. Note that the numerics breaks down close to the singularities. The solution corresponds to initial values $\vecX(0) = (6, 5, 7)$, $\dot{\vecX}(0) = (6,7,5)$ and $J^+ = 8$. We choose $\mu = 7$ and $\ell = 1$\label{fig:2C}.} 
\end{center}
\end{figure}

\begin{figure}[H]
\begin{center}
{\bf Soliton with analytic centre --- Taylor expansion}\\
\includegraphics[height=180mm]{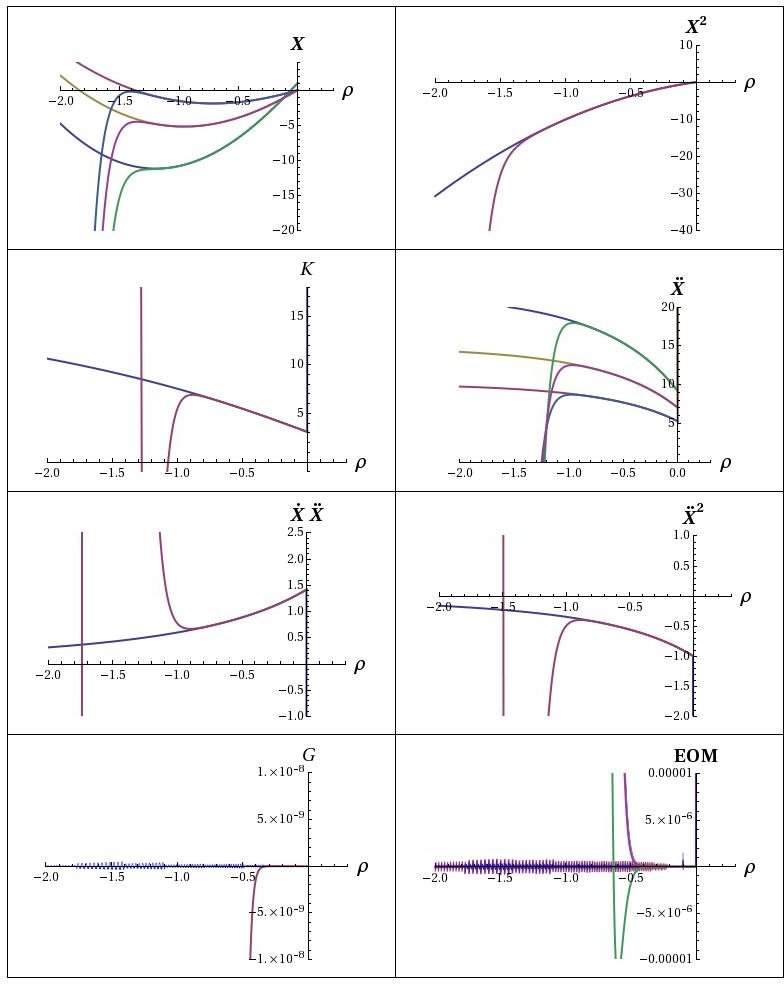}
\caption{The figure shows a solution in the generic sector with an analytic centre at $\rho=0$. In the plots both the Taylor expansion and the corresponding extension by algorithm 3 are shown. We have chosen $\mu = 5$ and $\ell = 1$, and the solution corresponds to thee upper sign in \eqref{eq:as48} and $a=3.7$ which is in between the Einstein and warped values. \label{fig:TnE}}
\end{center}

\end{figure}
\begin{figure}[H]
\begin{center}
{\bf Soliton with analytic centre --- Match and evolve}\\
\includegraphics[height=180mm]{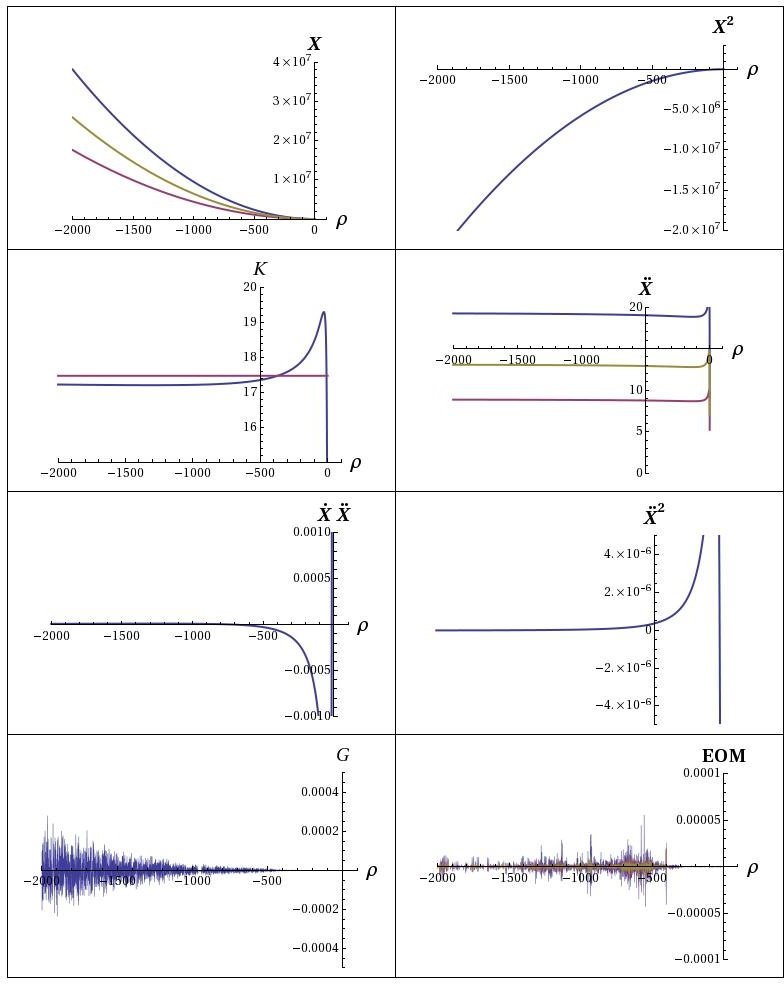}
\caption{The same solution as in Fig.~\ref{fig:TnE}, but with a wider range of $\rho$. Asymptotically the solution looks like warped AdS. The straight line in the plot of $K$ is the exact value for warped AdS. It is clear from the plots of $G$ and ${\bf EOM}$ that the numerics are problematic for large negative $\rho$.\label{fig:AnalCenters}}
\end{center}
\end{figure}

\end{document}